\documentclass[12pt]{article}
\usepackage[utf8]{inputenc}

\usepackage{hyperref}
\usepackage{amsmath}
\usepackage{amsfonts}
\usepackage{ulem}
\usepackage{graphicx}
\usepackage{latexsym}
\usepackage{slashed}
\usepackage{authblk}
\usepackage{fullpage}

\newcommand{\eq}{\begin{equation}}
\newcommand{\eqx}{\end{equation}}
\newcommand{\eqn}{\begin{eqnarray}}
\newcommand{\eqnx}{\end{eqnarray}}
\newcommand{\be}{\begin{equation}}
\newcommand{\ee}{\end{equation}}
\newcommand{\bea}{\begin{eqnarray}}
\newcommand{\eea}{\end{eqnarray}}

\newcommand{\f}[2]{\frac{#1}{#2}}

\newcommand{\LL}{{\mathcal L}}

\newcommand{\eps}{\varepsilon}
\newcommand{\al}{\alpha}

\newcommand{\dl}{\delta}
\renewcommand{\th}{\theta}

\newcommand{\bt}{\beta}
\newcommand{\gm}{\gamma}
\newcommand{\Gm}{\Gamma}

\newcommand{\Sg}{\Sigma}
\newcommand{\SSg}{\mathbf{\Sg}}

\newcommand{\OO}[1]{\mathcal{O}\left(#1\right)}

\newcommand{\qq}{\quad\quad}
\newcommand{\qqqq}{\quad\quad\quad\quad}

\newcommand{\non}{\nonumber \\}
\newcommand{\CR}{\non\cr}
\newcommand{\FGA}{\mathcal{A}}
\newcommand{\FGB}{\mathcal{B}}
\newcommand{\FGC}{\mathcal{C}}
\newcommand{\FGD}{\mathcal{D}}

\begin{document}

\normalem

\title{A simple description of holographic domain walls in confining theories -- extended hydrodynamics}

\author[a]{Romuald A. Janik\thanks{e-mail: \texttt{romuald.janik@gmail.com}}}
\author[b,c]{Matti J\"arvinen\thanks{e-mail: \texttt{matti.jarvinen@apctp.org}}} 
\author[d]{Jacob Sonnenschein\thanks{e-mail: \texttt{cobi@tauex.tau.ac.il}}} 

\affil[a]{\emph{Institute of Theoretical Physics, Jagiellonian University, \L ojasiewicza~11, 30-348~Kraków,~Poland.}}
\affil[b]{\emph{Asia Pacific Center for Theoretical Physics, Pohang 37673, Republic of Korea}}
\affil[c]{\emph{Department of Physics, Pohang University of Science and Technology, Pohang 37673, Republic of Korea}}
\affil[d]{\emph{The Raymond and Beverly Sackler School of Physics and Astronomy}\\
	\emph{Tel~Aviv University, Ramat Aviv 69978, Israel}}

\hfill{APCTP Pre2021 - 011}

{\let\newpage\relax\maketitle}

\begin{abstract}
In the context of theories with a first order phase transition, we propose a general covariant description of coexisting phases separated by domain walls using an additional order parameter-like degree of freedom. In the case of a holographic Witten model with a confining and deconfined phase, the resulting model extends hydrodynamics and has a simple formulation in terms of a spacetime action with corresponding expressions for the energy-momentum tensor.
The proposed description leads to simple analytic profiles of domain walls, including expressions for surface tension density, which agree nicely with holographic numerical solutions, despite the apparent complexity of those gravitational backgrounds. 
\end{abstract}

\newpage
\section{Introduction}

The  seminal transition from a deconfined  phase to a confining one is one of the most important processes in elementary particle physics. 
It is of relevance both for the physics of the early universe and for the process of transforming quark-gluon plasma into the observed hadrons in the course of heavy-ion collisions.
In addition to the observational and experimental motivations,  also from a theoretical point of view the study  of the deconfined to confining transition   is   a 
fundamental and 
poorly  understood building block  of QCD. 

Some time after the discovery of the holographic duality, it has been realized that the gravitational backgrounds  may serve as useful  laboratories to study separately  confining and deconfined dynamics and then also  the transition between them. 
A variety  of  holographic backgrounds have been proposed for that purpose; some of them are bottom-up models and others are top-down  ones. 

In the early years of the AdS/CFT correspondence, phase transitions were primarily studied in the equilibrium setting by comparing the free energies of two distinct dual gravitational backgrounds. The issue of describing a passage through the phase transition in real time has only been addressed quite recently in a number of works, starting from a linearized analysis~\cite{Janik:2015iry, Janik:2016btb} to fully dynamical studies of phase separation and domain wall dynamics~\cite{Gursoy:2016ggq, Attems:2017ezz, JJS, Attems:2018gou, Attems:2019yqn, Bellantuono:2019wbn, Attems:2020qkg, Ecker:2021ukv}.

These studies could deal successfully primarily with cases where both phases of the relevant theories were deconfined. The full gravitational description in the most physically interesting case of a transition between a confining and a deconfined phase has proved so far to be prohibitively difficult, apart from the pioneering work~\cite{Bantilan:2020pay} dealing with the time evolution of plasma-balls. The goal of the present paper is to propose a simple description of the coexisting phases and domain walls from the boundary field theory point of view, which could serve potentially as a proxy for the full gravitational description in the same sense as hydrodynamics corresponds to dual gravitational backgrounds through the fluid/gravity duality~\cite{Bhattacharyya:2008jc}.
We believe, however, that the range of applicability of the proposed description is in fact very general and goes beyond holography.

Let us mention here two lines of research, whose scope partially overlaps with our approach. 

In~\cite{Attems:2017ezz}, a boundary description of evolving domain walls in theories with two deconfined phases was proposed in terms of second order ``spatial'' hydrodynamics.
Since our main focus of interest are theories with the confining/deconfined transition, a purely hydrodynamic description would not be applicable. Moreover, we would like to obtain analytically the very simple looking domain wall profiles which are seen in a wide variety of holographic models.

Reference \cite{Bigazzi:2020phm} studied Linde's approach to tunneling at finite temperature~\cite{Linde} in a theory with the confining/deconfined transition through a certain specific Ansatz on the gravitational side. Our approach is, in contrast, based on boundary field theory considerations and is \emph{a-priori} applicable to very general dynamical configurations including hydrodynamic flows on the deconfined side.
There are, nevertheless, marked qualitative similarities as a scalar field with a double well kind of potential appears naturally both here and in~\cite{Bigazzi:2020phm} (see also the earlier work~\cite{Creminelli:2001th}).
 
Our present work is based on a prototype model utilizing the idea of \cite{Witten:1998zw} of compactifying  the  Euclidean time coordinate and one space coordinate on an $S^1$. The analysis of this model  made in \cite{Witten:1998zw} and \cite{Aharony:2006da}  yielded the holographic confining/deconfined phase diagram.
The two dimensional sub-manifold of the full geometrical background that is spanned by the compact coordinate    and the holographic coordinate,   can have a topology of either  cylinder (cy.) or a cigar (ci.). 
It is well known that for a ci. topology that involves the Euclidean time  direction  the background is that of a black-hole and it is the dual of the thermal deconfined phase. On the other hand the ci. that involves a space coordinate corresponds to a confining phase~\cite{Kinar:1998vq}. 
For backgrounds that depend only on the holographic coordinate there are only two non-trivial and non-singular possibilities that are depicted in figure~\ref{figphases}.


A numerical holographic study of a static domain wall interpolating between backgrounds that correspond to  a  confining  and deconfined  phases at $T=T_c$ was made in~\cite{Aharony:2005bm}. In principle a similar type of analysis could be done also for a dynamical domain wall at $T\neq T_c$.
  However, the gravitational description of such a composite system is extremely complicated.  Because of that we adopted  in this paper a  strategy of analyzing the system  at the level of  the energy-momentum tensor of the boundary field theory, and use the known numerical plasma-ball domain wall solution~\cite{Aharony:2005bm} as additional holographic input which  constrains both the overall structure and the precise coefficients of the field theory energy-momentum tensor.
 The main observation from  the analysis of the data from \cite{Aharony:2005bm} is that, in particular, the energy density throughout the domain wall can be fitted to a good precision with a simple Ansatz involving the hyperbolic tangent function. 

The guideline  idea of this paper  is to write down the energy momentum tensor as
 a linear combination of the energy-momentum tensors of the two coexisting confining and deconfined  phases and of the domain wall. 
The linear combination is controlled by a scalar field $\gamma(x,t)$ which in the static case takes the form of a hyperbolic $tanh$ function of $x$ the coordinate perpedicular to the domain wall. We identified two options for the interpolation and the dependence of the surface tension on  $\gamma$. We  find  a very good  fit  with the numerical results for both of these two options. 

We determine an action for $\gamma$. The solution of the corresponding equations of motion  for the static case  is the hyperbolic $tanh$ function mentioned above. For the non-static case of $T\neq T_c$ we couple the action of $\gamma$ to a hydrodynamical action so that the full action is  built from the  ``fields" $\gamma$ and $T$ the temperature. It turned out that   the coupling to  the hydrodynamics  rendered the symmetric  potential  of $\gamma$  with two minima into an asymmetric one.  For a small deviation from the critical temperature we solve the equations of motion and determine an accelerating domain wall solution. 
 

We  then present  several applications based only on   the general structure of the proposed energy-momentum tensor:  
(i) First we link its parametrization to  the surface tension of the domain wall.  (ii) Then we compare equations for an equilibrium circular droplet with standard thermodynamic considerations.
(iii) Finally we reproduce the formula for thermodynamic nucleation probability from \cite{Landau} using Euclidean solutions as suggested by Linde~\cite{Linde}.

The description proposed in the present paper was developed based on a conformal version of the holographic  model~\cite{Witten:1998zw} for which there exists the numerical solution of the domain wall~\cite{Aharony:2005bm}.   We believe, however, that the resulting framework should have a much more general range of applicability.

Indeed quite different (nonconformal) holographic models, which have a $1^{st}$ order phase transition between two \emph{deconfined} phases exhibit a domain wall profile which is very well fitted by a $tanh$ function~\cite{Attems:2019yqn,Attems:2020qkg}.
Since an underlying $tanh$ shape of the domain wall arises analytically in our framework, this suggests that the framework could also be applied in those settings. We verified this explicitly in the case of the nonconformal model of~\cite{JJS} also with two \emph{deconfined} phases.

Let us mention finally an interesting interpretation of our proposed framework.
Extracting  the free energy as a quartic polynomial in $\gm$
  essentially provides a Landau type description of a $1^{st}$ order phase transition. Our formulation (\ref{e.overallactionB}), in one of the two options, can be then understood as promoting the Landau order parameter $\gm$ to a dynamical effective field and coupling it in a natural way with hydrodynamic degrees of freedom.
In this sense our approach bears some similarity that of~\cite{Hindmarsh:2013xza}, for example, where the coupling of Landau-like order parameter to hydrodynamics was also used to study the dynamics of a $1^{st}$ order phase transition. In our approach, however, the hydrodynamics and the order parameter are degrees of freedom of a single theory, and only the deconfined phase is described through hydrodynamics.

The study of the holographic transition between confining and deconfined phases and of plasma balls  was presented  in many other publications. In addition to the ones mentioned earlier, a partial list includes  \cite{Shuryak:2005ia, Mateos:2006nu, Kobayashi:2006sb, Mahato:2007zm, BallonBayona:2007vp,Erdmenger:2007cm} and references therein, \cite{Gursoy:2008za, Bhattacharya:2009gm, Emparan:2009dj, Evans:2010iy, CasalderreySolana:2011us, Mandal:2011ws, Alho:2012mh, Figueras:2014lka, Dudal:2015wfn, Armas:2015ssd, Chesler:2016ceu, Gursoy:2017wzz, Arefeva:2018hyo}.   Recently, applications of holography to first order phase transitions in the early universe and to the production of gravitational waves has received a lot of attention~\cite{Bigazzi:2020avc,Ares:2020lbt,Bea:2021zsu,Bigazzi:2021fmq}.

The paper is organized as follows. After the introduction at section    \S 1, we review in section \S 2 the basic holographic setups that correspond to a confining phase and  to a  deconfining one. We use the prototype of Witten's model with a $(ci,cy)$ geometry for the confining phase and  $(cy,ci)$ geometry for the deconfined phase.
We spell out the geometrical parameterization and its corresponding parameters of the large $N_c$ gauge theory 
for both the low and high temperature phases. We then review the computation of the difference of the classical gravity action between the two phases that translates to the difference of the free energies in boundary field theory, and present a sketch of the phase diagram.  Section \S 3 is devoted to the plasma-ball domain wall. We review the numerical solution of  the domain wall~\cite{Aharony:2005bm}  that interpolates between the confining and deconfined phases. 
The structure  
of the domain wall  is then analyzed in \S 3.2. 
We discuss some observations on how this structure arises in gauge/gravity duality in \S 3.3. 
The next section, section \S 4 is devoted to the covariant description of the domain wall from the point of view of the boundary field theory. The general structure of the energy-momentum tensors is discussed in \S 4.1 and in \S 4.2: we address the comparison  of the   energy-momentum tensors  
with the holographic 
results 
which can be done by using two different options. We denote them as option A and option B. We compare the solution  with the numerical solution and analyze  the equation of motion for $\gm$.
\S 5 is devoted to an action density   formulation of the system.
We start with an action for perfect fluid hydrodynamics in \S 5.1, and then in \S 5.2 we present the action for the scalar field $\gm$ and finally in \S 5.3 we write down the final formulation. In \S 6 we discuss  some generic physical applications. The domain wall surface tension is described in  \S 6.1. Next we analyze the  circular droplet in \S 6.2. Thermodynamic nucleation probability and Euclidean solutions is determined in \S 6.3. For $T\neq T_c$ the domain wall is not static. The accelerating planar domain wall solution is written down in \S 7. This includes the effective actions and the corresponding accelerating solutions
in section \S 7.1 and  the coupling to hydrodynamics in \S 7.2. Comments on generality are made in \S 8  and in \S 9 we summarize, conclude and write down several open questions.
In appendix A we generalize the construction of the domain wall also to a six-dimensional setup.
In appendix~\ref{s.nonconformal} we checked that indeed our description works quite well also for domain walls in the non-conformal bottom-up holographic model of~\cite{JJS} with two deconfined phases for which we had numerical data.
Finally in appendix~C we derive the $n^\mu n^\nu$  terms in the energy-momentum tensor in the Witten model from an action.

\section{The Basic holographic setup}
\label{s.setup}

 To describe  using holography the transition between the confining  and deconfined  phases of the YM theory, one needs a holographic background that at low temperature is dual to the former phase and at high temperature to the latter.  The distinction between these two phases can be made by the expectation value of a rectangular  Wilson line that stretches along the time and one space directions: the confining and deconfined phases are  characterized  by  an area and  perimeter  law behaviours of the expectation value of such  Wilson lines respectively. In \cite{Kinar:1998vq}  the holographic  stringy duals of Wilson lines were analyzed and sufficient conditions for  confining background were identified. 
These criteria apply both to 
bottom-up and top-down confining backgrounds. In this work we will be interested mainly  in the latter but our conclusions will apply also to the former scenarios. Several    different gravitational backgrounds  that are confining backgrounds were written down (for a review see \cite{Erdmenger:2007cm} and references therein). 
It is well known that the gravitational duals of deconfined systems take the form of  black-hole backgrounds. 

Thus, the idea is to determine  a background that interpolates between these two types of backgrounds. At temperature equal to the critical temperature a static domain wall background is anticipated. This case was analyzed  thoroughly in \cite{Aharony:2005bm} for a background that asymptotes to $AdS_5\times S^5$ at the boundary.   At temperatures that are below or above the critical temperature the system should take the form of a dynamical domain wall.  

Originally, the gauge/gravity correspondence  dealt with supersymmetric systems. Various mechanisms have been incorporated to break supersymmetry.
  A prototype mechanism of supersymmetry breaking  is based on   compactifying one space-time coordinate   on $S^1$  and imposing  anti-periodic boundary conditions on fermionic fields~\cite{Witten:1998zw}. When the compactified coordinate is  the Euclidean time it  corresponds   to introducing temperature on the dual boundary field theory. Alternatively one can  compactify   one space-coordinate.  In this case, 
 in the limit of zero radius of compactification  the dimension of the boundary field theory is reduced by one.

	The two dimensional sub-manifold of the full geometrical background that is spanned by the compact coordinate denoted by either $\tau$ for the Euclidean time direction or $\phi$ for the space direction  and the holographic coordinate, denoted by $u$ can have a topology of either  cylinder (cy.) or a cigar (ci.). 
It is well known that for a ci. topology that involves the $\tau$ direction  the background is that of a black-hole.
Using the criteria for  a confining background mentioned above, it is easy to realize that  a ci. background with  $\tau$ direction is not dual to a  confining phase. On the other hand a ci. background along the $\phi$ direction does admit confinement.    

\begin{figure}[t!]
\begin{center}
\vspace{3ex}
\includegraphics[width=.80\textwidth]{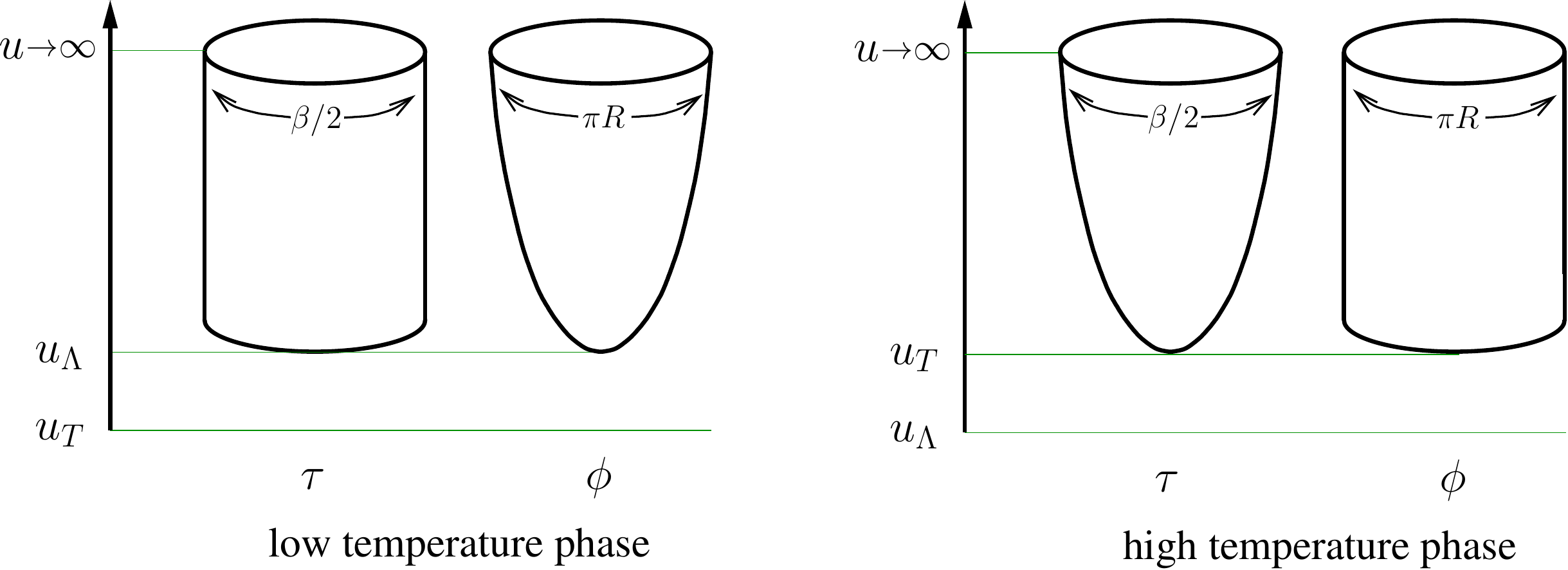}
\end{center}
\caption{ The topology of the solutions which dominate in the 
  low temperature (confined) and high temperature (deconfined) phases
  of the compactified D3 brane model, as reflected in the $(\tau, u)$ and $( \phi, u)$
  submanifolds.
  \label{figphases}}
\end{figure}

It turns out that for backgrounds  that correspond to the near horizon limit of large number of $D_p$ branes with one time and one space directions compactified there are solutions to the corresponding Einstein equations of motion that depend only on the holographic coordinate $u$ with the following topologies 
\be
 ( (\tau,u),(\phi,u)) = (cy, cy) \qquad \mathrm{or} \ \  (cy, ci)\qquad \mathrm{or}\ \  (ci, cy)\qquad \mathrm{or}\ \ (ci,ci) \ .
\ee
  The first case $(cy, cy)$ is flat and  the last  $(ci,ci)$ is necessarily singular.\footnote{Note, however, that a metric that depends not only on $u$ but also on other coordinates can be associated with a non-singular $(ci,ci)$ topology.}
Thus we are left with only two setups which are non-trivial and non-singular: the  $(ci, cy)$ and $(cy, ci)$  cases  that are depicted in figure~\ref{figphases} where we denote by $\beta$ and $2\pi R$ the circumference of the compact time $\tau$  and space $\phi$  directions respectively.

The analysis of a  holographic backgrounds that admit  this structure was performed in~\cite{Aharony:2006da}.
To  understand in detail the  properties of the geometrical background and its dual boundary field theory,  we briefly review the case of $D3$ branes.
We will refer to this model as a $d=3$ Witten model (with $d$ being the number of spatial dimensions or equivalently the noncompact space-time dimensions), when we want to emphasize its dimensionality or contrast with the original one~\cite{Witten:1998zw} with a nontrivial dilaton profile.
The first setup is with a cylinder in the $(\tau,u)$ sub-manifold and a cigar r in the $(\phi,u)$, namely the  $(cy,ci)$ set up. We refer to it as the low temperature setup. The boundary field theory dual to this background  in the limit of $R\rightarrow 0$ relates to 3d pure YM-like theory.
In \cite{Aharony:2006da} the background of large number of $D4$ branes was described~\cite{Brandhuber:1998er}. This corresponds in the same limit to 4d YM-like theory. The reason we take the case of $D3$ branes is that it is conformal and hence somewhat easier to handle. 
We then describe also the high temperature setup of the $(ci,cy)$ topology.

\subsection{ The (cylinder,cigar) low temperature  phase}
 
The $(cy,ci)$ background,
which  is a solution of the   type IIB string theory background
is characterized by the metric, the
RR four-form and a dilaton given by 
\bea\label{SSmodel} ds^2&=&\left(
\frac{u}{R_{D3}} \right)^{2}\left [ d\tau  ^2 +\delta_{ij}dx^i dx^j +
f(u) d\phi^2 \right ] +\left( \frac{R_{D3}}{u} \right)^{2} \left [
\frac{du^2}{f(u)} + u^2 d\Omega_5^2 \right ],
\CR
F_{(5)}&=& \frac{2\pi N_c}{V_5}\epsilon_5, \quad  
e^\phi = g_s , 
\quad
 R_{D3}^4 \equiv 4\pi g_s N_c l_s^4,\quad 
f(u)\equiv 1-\left( \frac{u_\Lambda}{u} \right)^4,
\eea   
where  $x^i$ ($i=1,2 $) are the
 uncompactified world volume coordinates of 
the D3 branes.   The volume of the five sphere $\Omega_5$ is
 denoted by $V_5$ and the corresponding volume form by $\epsilon_5$,
 $l_s$ is the string length and  $g_s$ is  
  the string coupling. As discussed above,  the sub-manifold of the background spanned by
 $\phi$ and $u$ has the topology of a cigar (as on the left-hand side of
 figure \ref{figphases}) where the minimum value of $u$ at
 the
 tip of the
 cigar is $u_\Lambda$. The tip of the cigar is non-singular if and
 only if the periodicity of $\phi$ is
\be\label{periodicity} \delta
 \phi = \frac{4\pi}{3}\left( \frac{R_{D3}^4}{u^2_\Lambda} \right)^{1/2} = 2\pi R
 \ee 
and we identify this with the periodicity of the circle that the
$3+1$-dimensional gauge theory lives on.


The parameters of this gauge theory, the four-dimensional gauge coupling
$g_4$, the low-energy three-dimensional 
 gauge coupling $g_3$ , the glueball mass scale $M_{gb}$, and the
 string tension $T_{st}$ are determined from the gravity background in
 the following form :
\bea\label{stringauge}
g_4^2&=&(2\pi) g_s ,\qquad 
g^2_{3}=\frac{g_4^2}{2\pi R}=
\left (\sqrt{\frac{g_s}{\pi N_c}} \frac{ u_\Lambda}{ l_s^2}\right )^{1/2},
 \qquad 
M_{gb} = \frac{1}{R}
\CR T_{st} &=& \frac{1}{2\pi
 l_s^2}\sqrt{g_{tt}g_{xx}}|_{u=u_\Lambda}= \frac{1}{2\pi l_s^2}\left(
 \frac{u_\Lambda}{R_{D3}} \right)^{2} =  \frac{\sqrt{4\pi N_c g_s}}{4 R^2} 
= \frac{\sqrt{\lambda_4}}{ R^2},\eea
where $\lambda_4 \equiv 2 g_4^2 N_c$, $M_{gb}$ is the typical scale of
the glueball masses computed from the spectrum of excitations around the background given\footnote{In fact, as was discussed in \cite{Sonnenschein:2015zaa}   part of the glueball spectra in  holography is described by closed strings in this background and not by the spectra of fluctuations of the bulk fields.} in 
(\ref{SSmodel}), and $T_{st}$ is the confining string tension in this
model (given by the tension of a fundamental string stretched at $u=u_{\Lambda}$
where its energy is minimized). The gravity approximation is valid
whenever $\lambda_4 \gg R$, otherwise the curvature at $u \sim
u_{\Lambda}$ becomes large. Note that as usual in gravity
approximations of confining gauge theories, the string tension is much
larger than the ``glueball"  mass scale in this limit determined for excitations of the bulk fields.  
As was mentioned above, the Wilson line of this gauge theory 
admits an area law behavior \cite{Brandhuber:1998er}, as can be easily seen using
the conditions for confinement of \cite{Kinar:1998vq}.

Naively, at energies lower than the Kaluza-Klein scale $1 / R$ the
 dual gauge theory is effectively three dimensional; however, it turns out
that the theory confines and develops a mass gap of order $M_{gb}=1/R$, so (in the
regime where the gravity approximation is valid) there
is no real separation between the confined three-dimensional fields and
the higher Kaluza-Klein modes on the circle. As discussed in \cite{Witten:1998zw},  
in the opposite limit of $\lambda_4 \ll R/l_s$, the theory approaches the
pure $2+1$ dimensional pure Yang-Mills theory at energies small compared to
$1/R$, since in this limit the scale of the mass gap is exponentially
small compared to $1/R$. It is believed that there is no phase transition
when varying $\lambda_4 l_s /R$ between the gravity regime and the pure
 Yang-Mills regime, but it is not clear how to check this.

\subsection{The (cigar,cylinder)  high temperature phase}

The high temperature  phase which  associates with the $(ci,cy)$ setup of figure~\ref{figphases}.
is described by the following gravity background  
\bea \label{actionhigh}
ds^2&=&\left(
\frac{u}{R_{D3}} \right)^{2}\left [f(u) d\tau^2+ \delta_{ij}dx^{i}dx^j
+ d\phi^2\right ] +\left( \frac{R_{D3}}{u} \right)^{2} \left [ u^2
d\Omega_5^2 + \frac{du^2}{f(u)} \right ],
\label{unflavmetr}\CR
F_{(5)}&=& \frac{2\pi N_c}{V_5}\epsilon_5, \quad  
e^\phi = g_s ,\quad
 R_{D3}^4 \equiv \pi g_s N_c l_s^4,\quad 
f(u)\equiv 1-\left( \frac{u_T}{u} \right)^4,
\eea
in the same notations as in (\ref{SSmodel}).

In a similar manner to the low temperature phase, now to avoid a singularity at the tip of the cigar, if we require that the periodicity for the $\tau$ coordinate obeys the same condition as in (\ref{periodicity}) but now with $u_T$ replacing $u_\Lambda$.  This replacement has to be made also in the relations between  $g_4^2$ and $g_3^2$ and the parameters of the background. 
An important difference between the low and high temperature phases is that now following \cite{Kinar:1998vq}
 the string tension vanishes since
 \be
 T_{st} = \frac{1}{2\pi
 l_s^2}\sqrt{g_{tt}g_{xx}}|_{u=u_T}= \frac{1}{2\pi l_s^2}\left(
 \frac{u_\Lambda}{R_{D3}} \right)^{2}\sqrt{1-\left( \frac{u_T}{u}\right)}_{u=u_T} =0 \ .
 \ee

\subsection{Free energy and the phase diagram}
The determination of the  holographic phase diagram for the model of \cite{Brandhuber:1998er}  was performed in~\cite{Aharony:2006da}. Here we summarise the results and convert them to Witten's of three dimensional YM theory~\cite{Witten:1998zw}. 
In order to decide which background dominates at a
given temperature $T$ one needs to compute the free energies of these
backgrounds, given  by the
classical supergravity action times the temperature, and see which one
has a lower free energy. As usual, the classical action actually
diverges and needs to be regulated and renormalized. This can be achieved    by
determining the difference between the actions of the two solutions,
which turns out to be finite. 
 Using  this method it is clear that the free
energies of the two solutions coincide when the (asymptotic)
circumferences of the two circles are equal, $\beta = 2 \pi R$, since
the two solutions are identical then except for a relabelling of the
coordinates. Thus, at this temperature $T_c = 1 / 2\pi R$ there is a
first order phase transition between the two backgrounds. The
transition is of first order since the solutions do not smoothly
connect there, but continue to exist as separate solutions both below
and above the transition. It is easy to see that the background 
 $(cy,ci)$  dominates at low
temperatures $T < 1 / 2 \pi R$, while the background $(ci,cy)$   dominates at high temperatures, $T > 1 / 2\pi R$.

The physical interpretation of this phase transition 
is straightforward.  As was mentioned above, the stringy dual of the Wilson line stretched between an external quark anti-quark pair located on the boundary admits an area low behaviour at low temperature and a perimeter law at the high temperature phase. 

Similarly, a computation of the value of the free energy in the
two phases (which requires adding appropriate counter-terms) yields a
result of order $N_c^0$ in the low-temperature phase, and a result of
order $N_c^2$ in the high-temperature phase.
\begin{figure}[t!]
\begin{center}
\vspace{3ex}
\includegraphics[width=.6\textwidth]{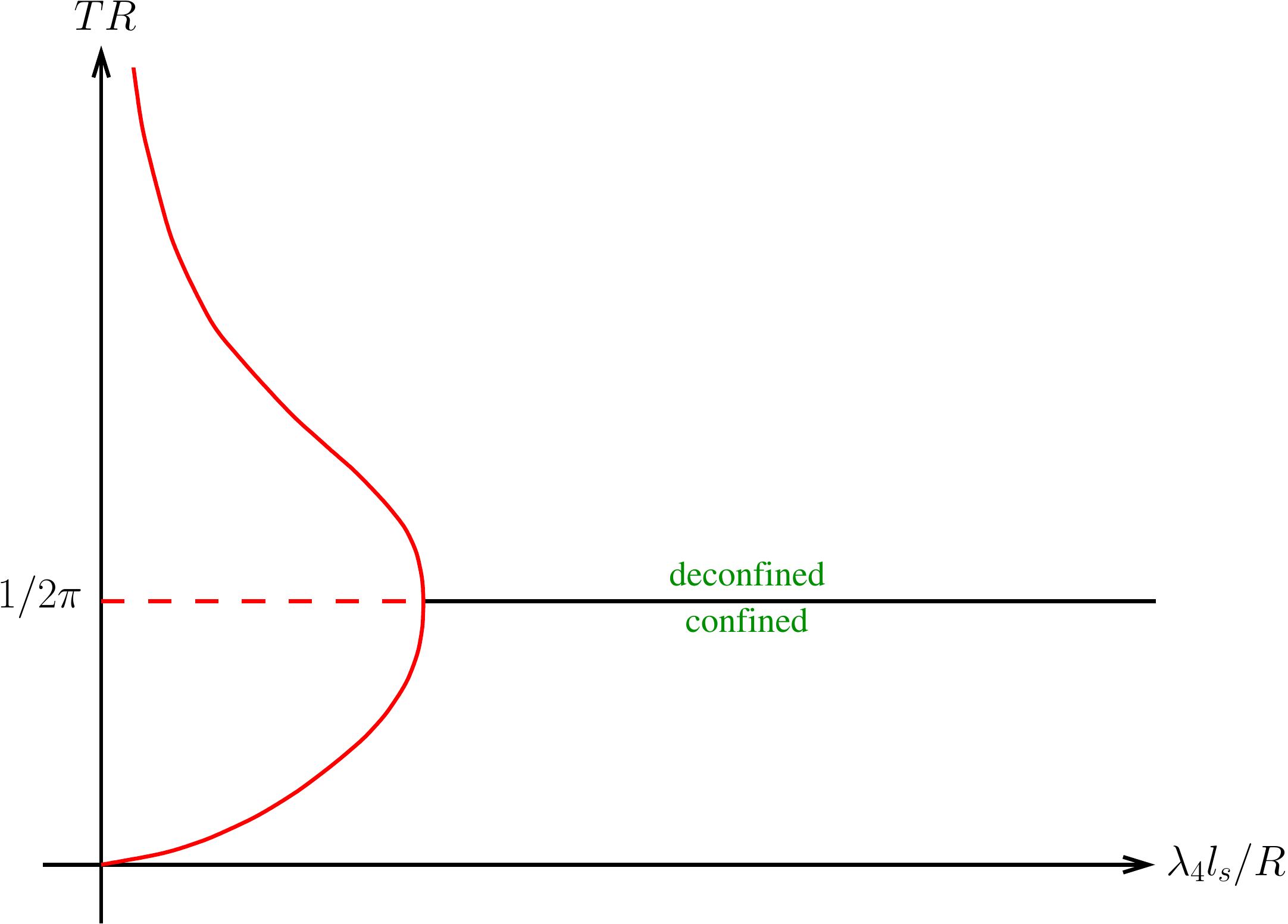}
\end{center}
\caption{Possible phase diagram of the compactified D3 brane model as a function of the gauge coupling $\lambda_4$ and the temperature. \label{phasediagram}}
\end{figure}
 
  As one increases $\frac{R}{l_s}$ compared to the scale set by
the four dimensional gauge coupling $\lambda_4$, the supergravity
background becomes highly curved, and the theory at low energies
approaches the three dimensional pure Yang-Mills theory. It is believed
  that the confining deconfined    transition
described above is connected to the 
transition of the
large $N_c$ pure Yang-Mills theory in this limit. However, this
connection is somewhat subtle \cite{Aharony:2005bm}, since even for large
$\frac{R}{l_s \lambda_4}$ it is clear by the symmetry arguments described above
that there is always a phase transition line at $T=1/2\pi R$ 
while the  confining deconfining
phase transition in the pure Yang-Mills theory occurs at a temperature
of order $\Lambda_{QCD} \ll 1/R$. Due to the $T \leftrightarrow 1/2\pi
R$ symmetry there must then be at least one more phase transition line
at a temperature much larger than $1/R$. Thus, it was argued in \cite{Aharony:2006da} that  new phases
must appear as one approaches the three dimensional Yang-Mills limit. The
phase structure of this theory in that limit includes several phase
transition lines, which, presumably, all join into the line $T = 1 /
2\pi R$ in the gravity limit. A minimal conjecture for the phase
structure is depicted in figure~\ref{phasediagram}.


\section{The plasma-ball domain wall}
\label{s.domainwall}

In this section we discuss various aspects of the domain wall between confining and deconfining phases from the holographic perspective. This will serve to motivate the extended hydrodynamic picture for the domain wall and coexisting phases which we introduce and study in the remaining part of the paper.

\subsection{Numerical study of the domain wall in gauge/gravity duality}
\label{s.domainwall_numerical}

The most important motivation for our work is the numerical study of the structure of the domain wall in gauge/gravity duality. The main observation is that, in particular, the energy density throughout the wall can be fitted to a good precision with a simple Ansatz involving the hyperbolic tangent function. We will study this in detail in the case of the $(cy, ci)$ to $(ci,cy)$ confinement-deconfinement transition, but the same has been shown to hold also for transitions between two (deconfined) black hole phases in five dimensional Einstein-dilaton gravity~\cite{Attems:2019yqn}.

In~\cite{Aharony:2005bm} the domain wall solution between a confining $(cy,ci)$ geometry and the deconfining $(ci,cy)$ geometry was constructed numerically by solving the Einstein equations for higher dimensional gravity. More precisely, the authors of this article considered planar domain wall solutions to the Einstein gravity with a cosmological constant in five and six dimensions, which are closely related to AdS$_5$ ($d=3$) and AdS$_6$ ($d=4$) spaces, respectively. Notice that the six dimensional geometry  differs from the holographic model~\cite{Witten:1998zw, Brandhuber:1998er} of compact  D4 background as there is no dilaton (or the dilaton is constant) and the background is conformal. 
We will discuss here the domain wall in this five dimensional geometry, and the results for the six dimensional geometry are given in Appendix~\ref{app:d4}. The numerical study is restricted to the critical temperature where the solution is static. Generalizations to spherical domains~\cite{Figueras:2014lka} and real-time evolution producing such spherical plasma balls~\cite{Bantilan:2020pay} have been also considered in the literature.

The confining five-dimensional geometry (AdS$_{5}$ with one spatial direction compactified) is the reduction of~\eqref{SSmodel} to five dimensions:
\be\label{eq:confgeom}
 ds^2_\mathrm{con.} = \left(\frac{u}{R_{D3}}\right)^2(d\tau^2 
 +\delta_{ij}dw^idw^j+f(u)d\phi^2) +\left(\frac{R_{D3}}{u}\right)^2\frac{du^2}{f(u)}
\ee
where 
\be
 f(u) = 1 - \left(\frac{\pi R_{D3} T_c}{u}\right)^{4} \ ,
\ee
$R_{D3}^2$ is the AdS radius, $\tau$ is the euclidean time, $\phi$ is the compactified spatial coordinate, the indices $i$, $j$ run from 1 to 2, and the critical temperature is related to the radius of compactification through $T_c = 1/2\pi R$. 
This solution and the deconfining geometry at critical temperature,
\be\label{eq:deconfgeom}
 ds^2_\mathrm{decon.} = \left(\frac{u}{R_{D3}}\right)^2(f(u)d\tau^2 +\delta_{ij}dw^idw^j +d\phi^2) +\left(\frac{R_{D3}}{u}\right)^2\frac{du^2}{f(u)u^2}
\ee
satisfy the five-dimensional Einstein equations with a cosmological constant,
\be
 R_{\mu\nu} = -\frac{4}{R_{D3}^2} g_{\mu\nu} \ .
\ee

The interpolating geometry between these two geometries was solved numerically by using a special set of coordinates (see~\cite{Aharony:2005bm} for details) but we find it useful to write it in Fefferman-Graham coordinates:
\be \label{eq:FGmetric}
ds^2 = \frac{1}{z^2}\left[\FGA(z,x)d\tau^2 + \FGB(z,x) d\phi^2 + \FGC(z,x)dy^2 + \FGD(z,x)dx^2 + dz^2\right] \ ,
\ee
where $y$ denotes the spatial coordinate parallel to the domain wall: $y = w_1$. The coordinates $z$ and $x$ can be understood to be nontrivial combinations of $w_{2}$ and $u$. In Fefferman-Graham coordinates, the holographic coordinate $z$ vanishes at the UV boundary (so that it is $z \sim 1/u$ near the boundary). The spatial coordinate $x$ is interpreted as the coordinate perpendicular to the wall. The factors $\FGA$, $\FGB$, $\FGC$, and $\FGD$ depend on the coordinates $z$ and $x$ only. The units of the solution (also setting $R_{D3}$ to one) are chosen such that %
\be
 \FGA\, , \ \FGB \, , \ \FGC\, , \ \FGD\ \to 1 \quad \mathrm{as} \quad z \to 0 \ .
\ee
After this the overall energy scale $\sim T_c$ is the only free parameter.
The solution enjoys a $\mathbb{Z}_2$ symmetry which reflects the symmetry of the setup under the exchange $\tau \leftrightarrow \phi$ at the critical temperature:
\be \label{eq:FGsymm}
 \FGA(z,x) = \FGB(z,-x) \ , \qquad \FGC(z,x)=\FGC(z,-x)\ , \qquad \FGD(z,x)=\FGD(z,-x) \ ,
\ee
where we chose the center of the domain wall to lie at  $x=0$.
In our conventions the solution asymptotes to the deconfined geometry~\eqref{eq:deconfgeom} (confined geometry~\eqref{eq:confgeom}) as $x \to -\infty$ ($x \to +\infty$). 

Before going to the structure of the domain wall in these solutions, we discuss the qualitative behavior and the topology of the interpolating geometry. In the bulk the geometry ends at a horizon at some $z=z_h(x)$, where
\begin{align}\label{eq:confcond}
 \FGA(z_h(x),x) &= 0 \ ,& \quad (x&\le 0)\ ; & \\ 
 \FGB(z_h(x),x) &= 0 \ ,& \quad (x&\ge 0)\ . &
\label{eq:deconfcond}
\end{align}
These conditions correspond to the end points of the cigars in the confining $(cy,ci)$ and deconfining $(ci,cy)$ geometries, respectively. In Fefferman-Graham coordinates these end points are actually second order zeroes. 
In the middle of the wall, $x=0$, both the conditions~\eqref{eq:confcond} and~\eqref{eq:deconfcond} are satisfied at some (finite) $z_h$. Therefore the $x=0$ slice of the interpolating solution has the topology $(ci,ci)$. 

\begin{figure}[t!]
\begin{center}
\includegraphics[width=.65\textwidth]{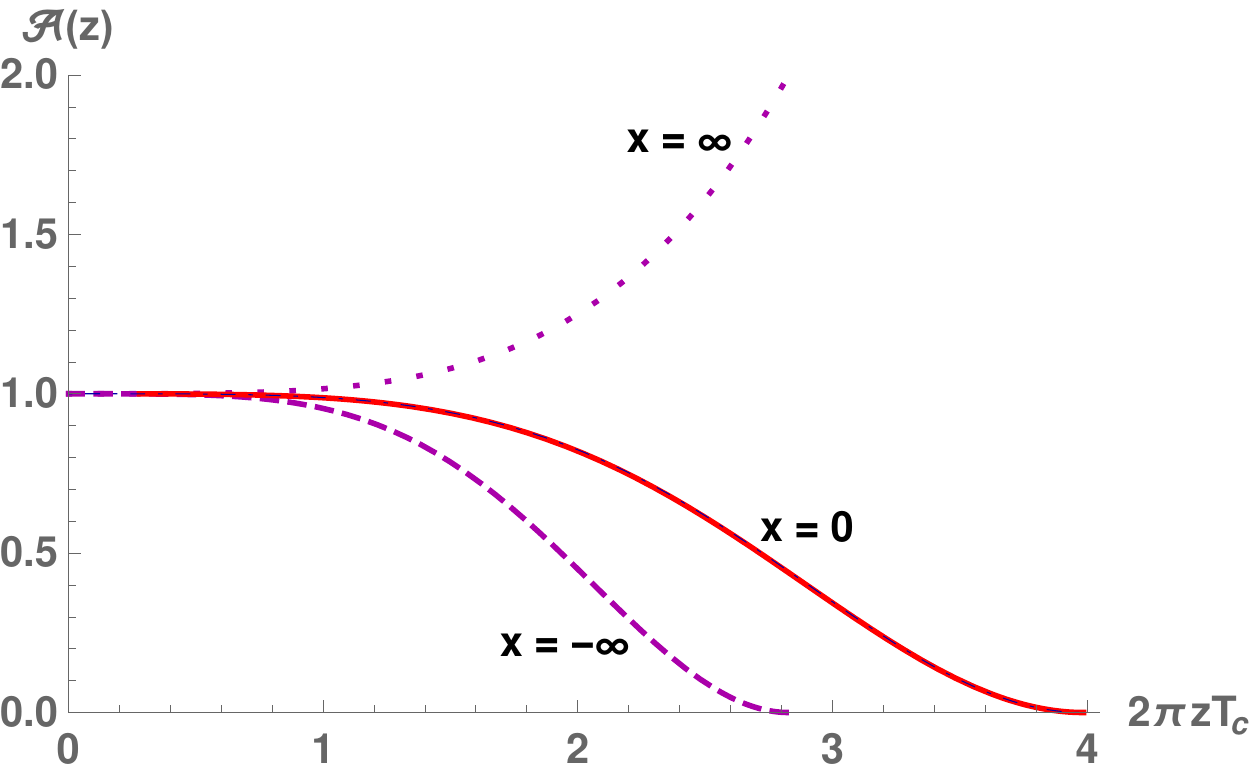}
\end{center}
\caption{Evolution of the metric component $\FGA(z) = z^2g_{\tau\tau}(z)$ in the Fefferman-Graham coordinates with varying $x$. The dashed and dotted magenta curves show the metric in the cigar (deconfined) and cylinder (confined) limits, respectively. The red curve is the metric in the middle of the domain wall, $x=0$. The dot-dashed thin blue curve is the cigar geometry with $z$-dependence rescaled by the factor $1.41$.
\label{fig:cigarevolution}}
\end{figure}

This observation may be verified by plotting the components of the metric: see figure~\ref{fig:cigarevolution} for the shape of $\FGA(z,x)$ in the middle as compared to the limiting functions\footnote{Notice that the ``cylinder'' form at $x \to +\infty$ is $\FGA(z)=1+z^4/z_h^4$ in the Fefferman-Graham coordinates, while for the flat $(cy,cy)$ geometry one has $\FGA(z)=1$.} at $x \to \pm\infty$. Due to the symmetry~\eqref{eq:FGsymm}, $\FGB(z,x)$ shows the same behavior (with $x=-\infty$ and $x=+\infty$ exchanged). We notice that the factors $\FGA(z,0)$ and $\FGB(z,0)$ in the middle of the wall (the red curve in the plot) are almost exactly given by a rescaled cigar geometry (the dot-dashed blue curve). The cigars in the middle are therefore elongated by a moderate factor $1.41$ with respect to the cigars in the confining and deconfining limits. This also means that the dependence of $z_h(x)$ is on $x$ is rather weak.

This result may appear surprising given the fact that the ($x$-independent) $(ci,ci)$ solution is singular. However if this topology only appears in a single slice of a higher dimensional geometry, there is no singularity. 

\begin{figure}[th]
\begin{center}
\includegraphics[width=0.9\textwidth]{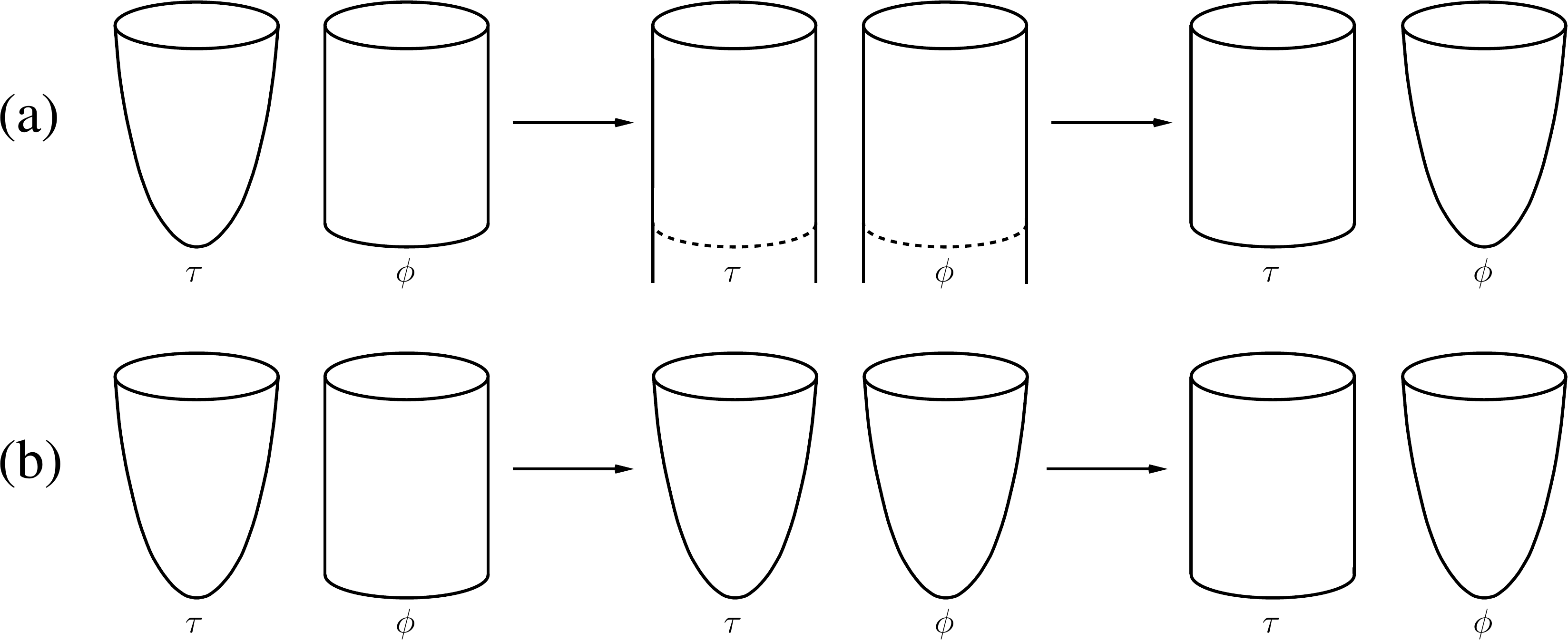}%
\end{center}
\caption{
Possible ``interpolations'' between the deconfined and confined geometries in figure~\protect\ref{figphases}. Top row (a): interpolation through the $(cy,cy)$ geometry. Bottom row (b): interpolation through the $(ci,ci)$ geometry.
\label{fig.cigarev}}
\end{figure}

This observation means roughly the following. There are two simple ``idealized'' ways to interpolate between the $(cy,ci)$ and $(ci,cy)$ geometries in~\eqref{eq:confgeom} and in~\eqref{eq:deconfgeom} (see the sketch in figure~\ref{fig.cigarev}). The first one is $(ci,cy)\to (cy,cy) \to (cy,ci)$ shown in figure~\ref{fig.cigarev}a: We start from the geometry~\eqref{eq:deconfgeom} and treat $T_c$  in $f(u)$ as a parameter. We smoothly reduce its value taking $T_c \to 0$ in the end which leads to the $(cy,cy)$ geometry with $f(u) \to 1$, i.e., the pure AdS$_{5}$ solution. We then introduce the blackening factor in the $d\phi^2$ component, turning back nonzero $T_c$ and smoothly evolve it to the original value, finally obtaining the $(cy,ci)$ solution.
The second one is $(ci,cy)\to (ci,ci) \to (cy,ci)$ shown in figure~\ref{fig.cigarev}b: Now we add instead a second blackening factor in the $d\phi^2$ term from the start, and smoothly evolve the parameter $T_c$ in this blackening factor to the value of the blackening factor of the $d\tau^2$ term, therefore obtaining a $(ci,ci)$ geometry. Then we reduce the value of the $T_c$ parameter of the blackening factor of the $d\tau^2$ term smoothly to zero, which gives the final  $(cy,ci)$ geometry. 
We stress that neither of these simple interpolations give rise to a solution of the five-dimensional Einstein equations, with the understanding that the interpolation is understood as a dependence on the coordinate $x$ perpendicular to the domain wall. However our point is that the interpolation defined by the $x$-dependence in the geometry of~\cite{Aharony:2005bm} 
which does solve the Einstein equations, is clearly closer to the second than the first option. The main difference between the numerical geometry and the ideal picture appears to be the fact that the cigars of the $(ci,ci)$ slice are elongated, as shown in figure~\ref{fig:cigarevolution}.

\begin{figure}[t!]
\begin{center}
\includegraphics[width=.65\textwidth]{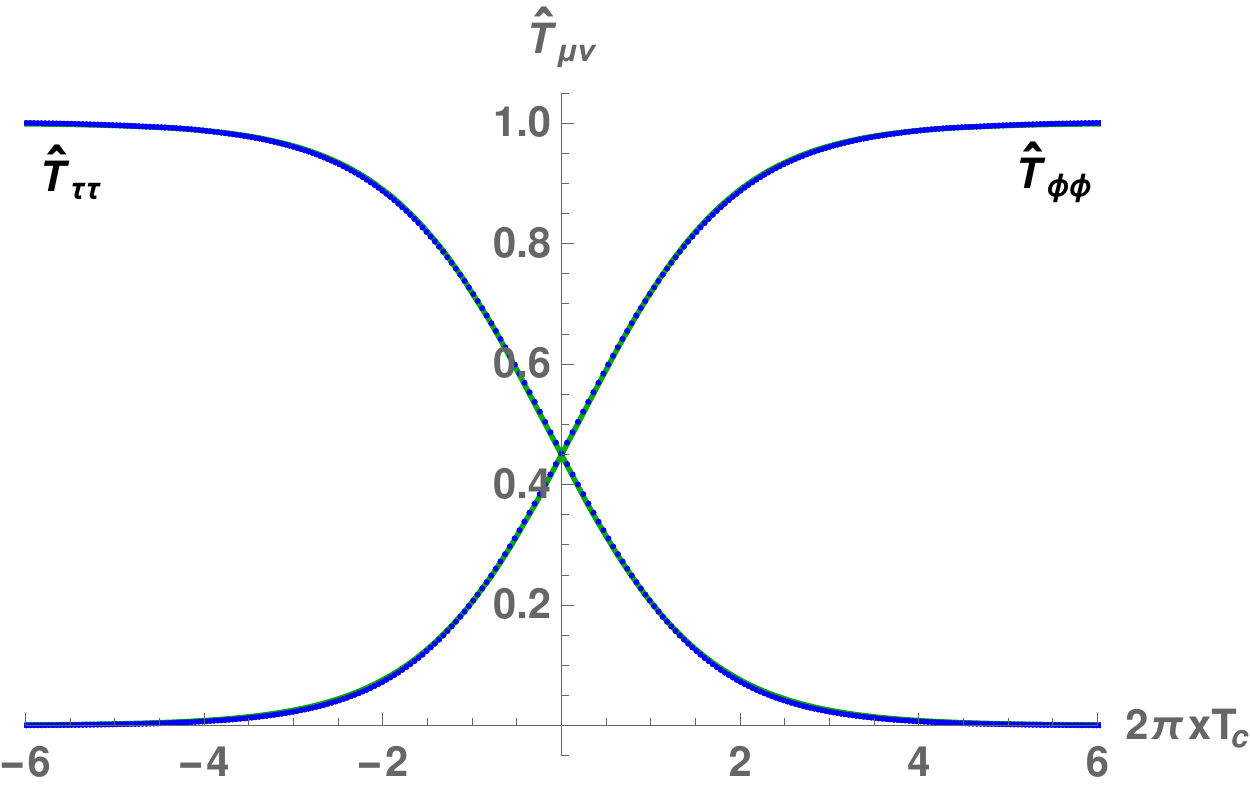}
\end{center}
\caption{Phenomenological fitting of the (normalized) components of the energy-momentum tensor $\hat T_{\tau\tau}$ and $\hat T_{\phi\phi}$. 
Blue dots show the energy-momentum tensor extracted from the numerically constructed geometry of~\protect\cite{Aharony:2005bm}, and the green curves are given by the fit in equation~\protect\eqref{eq:tanhfit}.
\label{fig:TttandTphiphi}}
\end{figure}

\subsection{Analysis of the energy momentum tensor of the domain wall}
\label{s.phenomenology}

\begin{figure}[t!]
\begin{center}
\includegraphics[width=.65\textwidth]{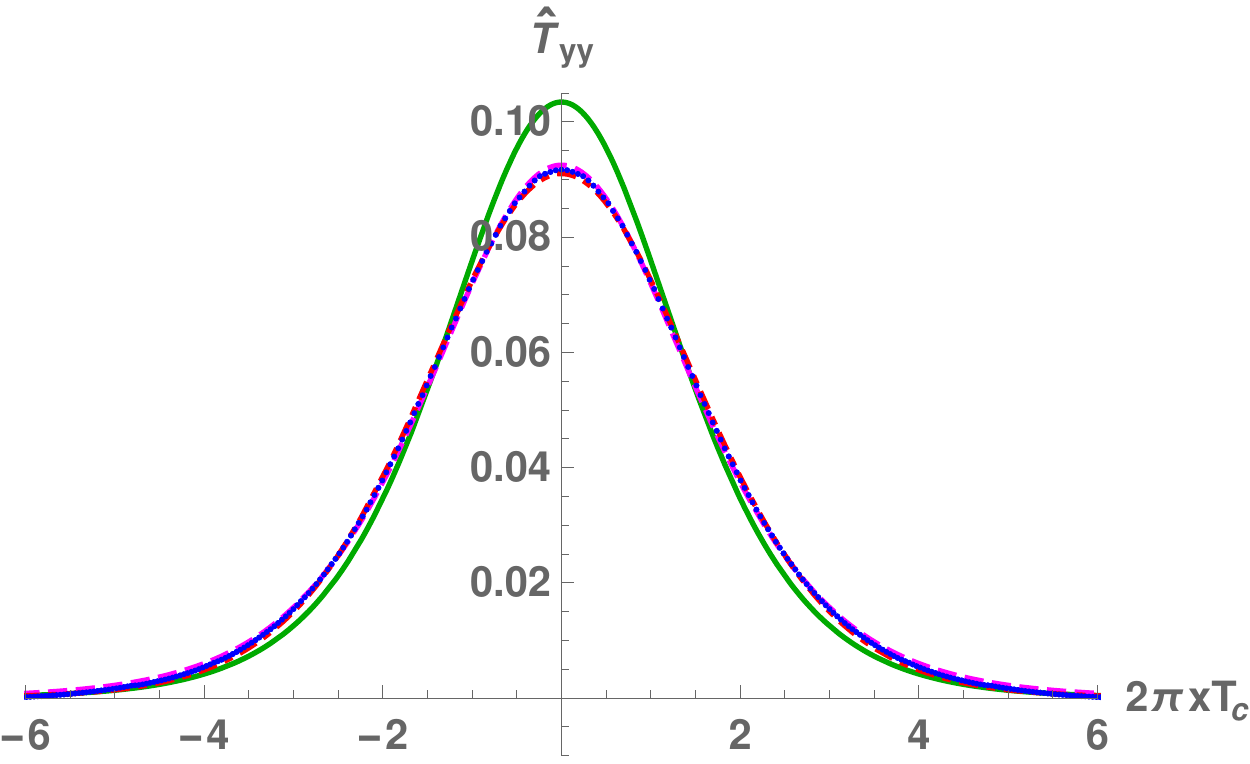}
\end{center}
\caption{Phenomenological fitting of the (normalized)  $\hat T_{yy}$ component of the energy-momentum tensor. 
The blue dots show the energy-momentum tensor extracted from the numerically constructed geometry of~\protect\cite{Aharony:2005bm}. The green curve is the prediction from the fit of figure~\protect\ref{fig:TttandTphiphi}, given in equation~\protect\eqref{eq:Tyypred}. The dashed magenta and dot-dashed red fits were obtained by using the former and latter formulas of equation~\protect\eqref{eq:coshfit}, respectively. \label{fig:Tyy}}
\end{figure}

We then go on discussing the field theory observables  
i.e., the components of the energy-momentum tensor $T_{\mu\nu}$.   They are found as the coefficients of the subleading $\mathcal{O}\left(z^{4}\right)$ terms of the geometry at the boundary. Notice that due to the form of the metric in~\eqref{eq:FGmetric}, $T_{\mu\nu}$ is diagonal. We have extracted these coefficients from the domain wall solution of~\cite{Aharony:2005bm} and show the results in Figs.~\ref{fig:TttandTphiphi} and~\ref{fig:Tyy}. The tensor is diagonal and only depends on the coordinate~$x$.  To plot the results, we use the normalized\footnote{Our sign conventions (in the Euclidean signature) are such that for the deconfined phase $T_{\tau\tau} = -\epsilon<0$ (where $\epsilon$ is the energy density) whereas the spatial component $T_{ii}$ of the energy momentum tensor are positive.} energy-momentum tensor
\be \label{eq:hatTmunudef}
 \hat T_{\mu\nu}(x) = -\frac{4 \pi G_{5}}{R_{D3}^3(\pi T_c)^{4}} \left(T_{\mu\nu}(x)-T_{xx}\right)
\ee
where $G_{5}$ is the Newton constant. That is, we divide out the trivial known proportionality constant, and subtract the value of $T_{xx}$ (the pressure perpendicular to the domain wall) which is constant for the domain wall solution due to the conservation law $\partial_\mu T^{\mu\nu} = 0$. The value of the constant is 
\be \label{eq:Txxvalue}
 T_{xx} = \frac{R_{D3}^3}{16\pi G_{5}}\left(\pi T_c\right)^{4} \ .
\ee
With this normalization, $\hat T_{\tau\tau} = 1$ for the deconfined geometry of~\eqref{eq:deconfcond}, for example.

The results for the normalized energy density $\hat T_{\tau\tau}$ and the pressure in the compactified direction $\hat T_{\phi\phi}$ are shown\footnote{Due to numerical inaccuracy the asymptotic values of $\hat T_{\mu\nu}$ agree with the analytic formulas in~\eqref{eq:hatTmunudef} and in~\eqref{eq:Txxvalue} only within an accuracy of about one or two per cent. Since we are mainly interested in the $x$-dependence at the domain wall, we ``renormalized'' the numerical data such that, for example, $\hat T_{\tau\tau}$ does approach one (zero) for $x \to -\infty$ ($x \to +\infty$).} in figure~\ref{fig:TttandTphiphi}.  
The blue dots show the data obtained from the numerical solution, and the green curves are given by the following fit:
\be \label{eq:tanhfit}
 \hat T_{\tau\tau}(-x) =  \hat T_{\phi\phi}(x) = \frac{1}{2} + \frac{1}{2} \tanh\left(q_* \frac{x-x_0}{2}\right) \ . 
\ee
The system is therefore in the deconfined (confined) phase in the domain $x<0$ ($x>0$). 
For the fit parameters we obtain
\be \label{eq:tfitresd3}
    \frac{q_*}{2\pi T_c}  \approx 1.148 \ , \qquad 2 \pi T_c x_0 \approx 0.181 \ .  %
\ee
The quality of the fit is very good: it appears to be even better than a similar fit for an interface between two deconfined phases in~\cite{Attems:2019yqn}.

Since $x^0$ is nonzero, there is a small but significant shift of the profile of the energy density from the middle of the domain wall, which is at $x=0$ as determined  by the $\mathbb{Z}_2$ symmetry. The shift also gives rise to a nontrivial component of the pressure parallel to the wall, $\hat T_{yy}$. This component is given by
\be \label{eq:Tyyrel}
 \hat T_{yy}(x) = 1-\hat T_{\tau\tau}(x)-\hat T_{\phi\phi}(x) \ ,
\ee
which is obtained by combining the tracelessness condition $T_\mu^\mu=0$ with~\eqref{eq:hatTmunudef}. Inserting here the fit from~\eqref{eq:tanhfit} gives the prediction
\be \label{eq:Tyypred}
 \hat T_{yy} = \frac{1}{2}\left[\tanh\left(q_* \frac{x+x_0}{2}\right)-\tanh\left(q_* \frac{x-x_0}{2}\right)\right] \ .
\ee
We compare this prediction (green curves), with the parameter values from~\eqref{eq:tfitresd3},  
to numerical data (blue dots) in figure~\eqref{fig:Tyy}. The maximal deviation is about ten percent and is due to two sources: the error in the $tanh$ fits and inaccuracy of the numerical solution (so that the trace $T_\mu^\mu$ is not precisely zero). The error from both sources has roughly the same size.  Notice that due to the relatively small size of the shift $x_0$, the magnitude of $\hat T_{yy}$ is suppressed roughly by a factor of ten with respect to the step size in figure~\ref{fig:TttandTphiphi}. This is why the errors are better visible in figure~\ref{fig:Tyy}.

We also fitted the data for $\hat T_{yy}$ directly using the following functions:
\be \label{eq:coshfit}
    \hat T_{yy}(x) = \frac{c^{(2)}}{\left(\cosh \frac{q_*^{(2)}x}{2}\right)^2} \ ,  \qquad
    \hat T_{yy}(x) = \frac{c^{(4)}}{\left(\cosh \frac{q_*^{(4)}x}{2}\right)^4} \ .
\ee
The first of these functions has an obvious motivation: taking the small $x_0$ approximation of~\eqref{eq:Tyypred} leads to
\be
 \hat T_{yy} \approx \frac{q_* x_0}{2} \frac{1}{\left(\cosh \frac{q_* x}{2}\right)^2}
\ee
and the motivation for the second function will become clear below. The results for the two parameter fits to data are shown in figure~\ref{fig:Tyy} with dashed magenta curves for the $cosh^{-2}$ fit and dot-dashed red curves for the $cosh^{-4}$ fit. The fits are so good that it is difficult to disentangle them for the data (blue dots) so we conclude that within the precision of our numerics, both fits work extremely well.
The fit parameters are in this case
\begin{align} \label{eq:cfitresd3}
    \frac{q_*^{(2)}}{2\pi T_c} & \approx 1.023 \ , \qquad c^{(2)} \approx 0.0925 \ ; &  \nonumber\\
    \frac{q_*^{(4)}}{2\pi T_c} & \approx 0.681 \ , \qquad c^{(4)} \approx 0.0910 \ . & 
\end{align}

From these fit results and the results for the $tanh$ fit in~\eqref{eq:tfitresd3} 
we observe that $q_*^{(2)}$ is close to $q_*$: 
we find $q_*^{(2)} < q_*$, and the difference is around ten per cent. We also notice that $q_*^{(2)}/q_*^{(4)} \simeq 1.5$. We will discuss these findings further by using an extended hydrodynamic framework below in section~\ref{sec.covdescript}.

\subsection{Modelling the domain wall in gauge/gravity duality}\label{s.toymodel}

The above findings, in particular the success of the simple $tanh$ fit in figure~\ref{fig:TttandTphiphi}, give rise to two natural questions. The first is if  there is a simple effective description (in terms of the field theory degrees of freedom) of the domain wall which gives rise to the observed structure. This is the main topic of the current article and will be addressed in detail in the following sections. The second is whether this structure can be obtained from gauge/gravity duality in some analytic approximation. This second question turns out to be difficult. We have however obtained some limited understanding on how the domain wall arises from the solutions of the Einstein equations, which we will now discuss.

We observe that the Einstein equations in Fefferman-Graham coordinates imply that there are two conserved bulk currents: the first ($J$) satisfies
\be \label{eq:Jdef}
 \partial_z J^z + \partial_x J^x = 0 \ , \qquad J^z = \frac{1}{z^3} \sqrt{\frac{\FGB\FGC}{\FGA}}\FGD^{3/2} \frac{\partial}{\partial z}\left( \frac{\FGA}{\FGD}\right) \ , \qquad  J^x = \frac{1}{z^3} \sqrt{\frac{\FGB}{\FGA\FGC}}\FGD^{3/2} \frac{\partial}{\partial x}\left( \frac{\FGA}{\FGD}\right) \ ,
\ee
and the second ($\tilde J$) is obtained by interchanging $\FGA \leftrightarrow \FGB$. The existence of such currents is no surprise: they reflect the Smarr relation for black holes which states that the energy of the black hole can be computed both from the asymptotics and from horizon data~\cite{Smarr:1972kt}. In the case of homogeneous black hole geometries in our setup, the conservation of the bulk current $J$ implies the standard thermodynamic relation $\eps + p = sT$. For the domain wall setup, we find that the conserved ``charge density'' is at the boundary
\be \label{eq:Jzbdry}
 J^{z}\big|_{z=0} \propto \hat T_{\tau\tau} -  \hat T_{yy}  
\ee
which shows a smooth step from one to zero over the domain wall as we see from Figs.~\ref{fig:TttandTphiphi} and~\ref{fig:Tyy}. At the horizon $z=z_h$ we find that
\begin{align}
     J^{z}\big|_{z=z_h} & = s_h \sqrt{2 \FGA''(z_h)} \ ,& &(x<0) \ ; &\qquad  
     J^{z}\big|_{z=z_h} & = 0 \ ,& &(x>0) \ ; & 
\end{align}
where
\be
 s_h = \frac{\sqrt{\FGB(z_h)\FGC(z_h)\FGD(z_h)}}{z_h^3}
\ee
is the area element of the black hole.

These results suggest the following picture. As we have seen in figure~\ref{fig:cigarevolution} the geometry in the deconfined phase varies relatively little as we move from $x=\infty$ to $x=0$. Therefore the bulk charge density $J^z(z,x)$ becomes essentially a step function at the horizon. During the evolution of the charge density from the horizon to the boundary, determined by equation~\eqref{eq:Jdef} and the other Einstein equations, the charge is diffused a bit which gives rise to the smooth step. Notice that because the total charge $\int dx J^z$ is conserved under this evolution  and because $J^x$ vanishes far from the domain wall, only minor rearrangement of the charge is possible.
 
We can make some quantitative estimates which support the above picture by resorting to simple modeling. First, because we have seen that $z_h$ depends only mildly on $x$ in the domain wall solution, we take $z_h$ to be constant. Second, we take the boundary condition for the current at the horizon to be exactly the step function. Third, we simply study the propagation due to a scalar field in plain AdS$_5$ geometry. Fourth, we look at the current $\Delta J=J-\tilde J = J - J(\FGA\leftrightarrow \FGB)$, which links to the difference $\hat T_{\tau\tau}-\hat T_{\phi\phi}$, instead of~\eqref{eq:Jzbdry}. This last simplification is useful because  $\hat T_{\tau\tau}-\hat T_{\phi\phi}$
 is odd in $x$ and has a simple $tanh$ profile for the domain wall solution without any shifts. The model setup may be formally obtained from~\eqref{eq:Jdef} (and its $\FGA \leftrightarrow \FGB$ interchanged counterpart) through the Ansatz $\FGC=1=\FGD$ and
\be \label{eq:modelAn}
 \FGA(z,x) = \exp(+\xi(z,x)) \ , \qquad  \FGB(z,x) = \exp(-\xi(z,x)) \ ,
\ee 
with $\xi(z,x)=-\xi(z,-x)$, so that the symmetry condition in~\eqref{eq:FGsymm} is satisfied. Since $\FGA$ and $\FGB$ approach one at the boundary, $\xi$ must vanish there.

\begin{figure}[t!]
\begin{center}
\includegraphics[width=.65\textwidth]{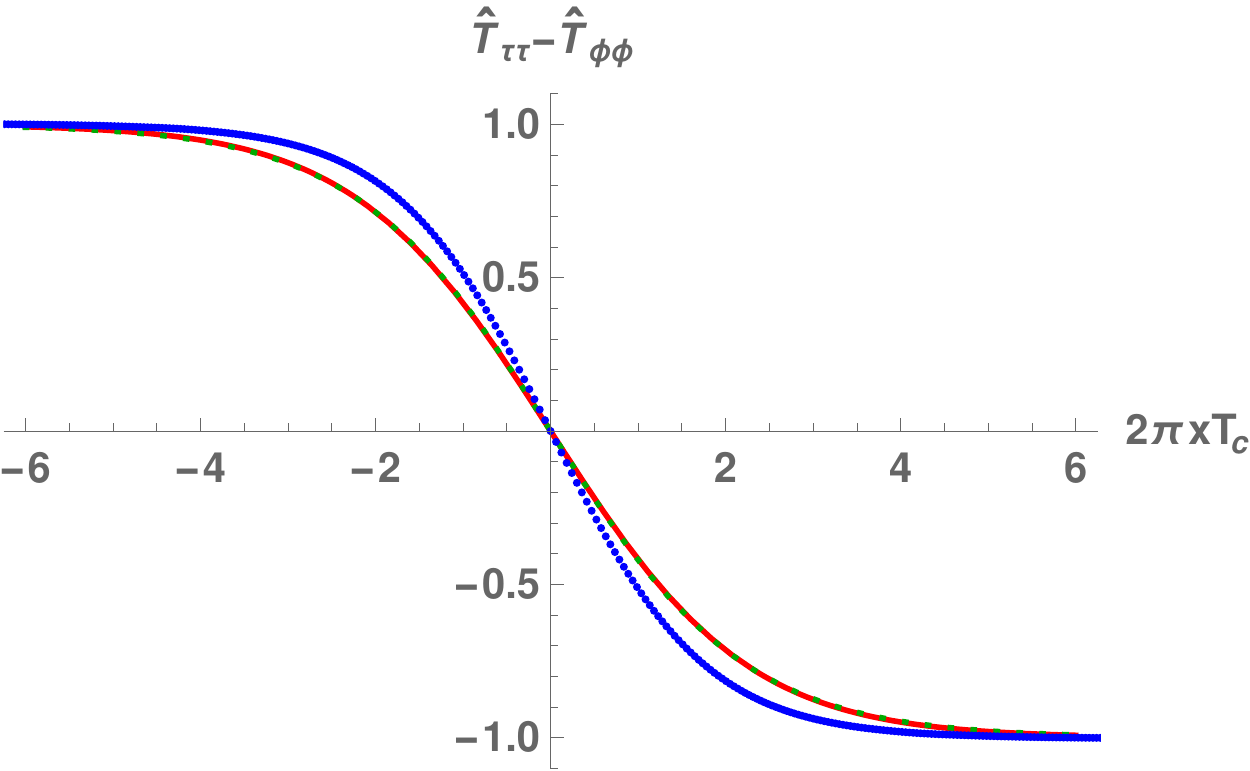}
\end{center}
\caption{$\hat T_{\tau\tau}-\hat T_{\phi\phi}$ from a simple model compared to the numerical data for the domain wall. The blue dots are numerical data from the exact solution, the red solid curve is given by~\protect\eqref{eq:DeltaJsol} at $z=0$, and the dotted green curve is a $tanh$ fit to the red curve. \label{fig:Tmunumodel}}
\end{figure}

We find that $\xi$ satisfies the Laplacian in AdS$_5$,
\be\label{eq:xiLap}
 z^3\frac{\partial}{\partial z} \left( \frac{1}{z^3} \frac{\partial}{\partial z} \xi(z,x)\right)  +  \frac{\partial^2}{\partial x^2} \xi(z,x) = 0 \ ,
\ee
and the current is given by 
\be \label{eq:Jxidef}
\Delta J^\mu  = \frac{2}{z^3} \frac{\partial}{\partial x_\mu} \xi(z,x) \ .
\ee  
The solution which satisfies the boundary conditions $\xi(0,x)=0$ and $\Delta J^z(z_h,x) = -\mathrm{sgn}(x)$ is given by
\be \label{eq:DeltaJsol}
 \Delta J^z(z,x) = -\int_{-\infty}^\infty dk\,\frac{ z_h I_1(k z) \sin (k x)}{\pi kz\, I_1(k z_h)} 
\ee
where $I_1$ is the modified Bessel function of the first kind.

Finally we compare the boundary value of $\Delta J^z$ to the numerical data for $\hat T_{\tau\tau}-\hat T_{\phi\phi}$ in figure~\ref{fig:Tmunumodel}. To do this, we fixed $z_h =1/(\pi T_c)$ as obtained from the AdS solution of~\eqref{eq:deconfgeom} with the understanding that $u = 1/z$. One might expect that we should use the Fefferman-Graham value for $z_h$ which is smaller by a factor of $\sqrt{2}$ but we feel that the above choice is better since the geometry of~\eqref{eq:deconfgeom} is closer to empty AdS$_5$ (which we are in practice using here) than the black hole geometry in Fefferman-Graham coordinates. Actually, the results for both values are roughly equally good, with the best fit value lying between the two. From the plot we see that $\Delta J^z(0,x)$ is very well fitted by the $tanh$ function (green dotted curve), and the value of $q_*/(2\pi T_c) \approx 0.897$ from this fit is slightly below the value $1.148$ obtained by fitting the data directly in~\eqref{eq:tfitresd3}.

We have therefore demonstrated that the simple model, with diffusion of the conserved bulk charge obtained from propagators for empty AdS$_5$, roughly agrees with the numerical data for the full solution of Einstein equations for the domain wall. In particular, the profile obtained for $\hat T_{\tau\tau}-\hat T_{\phi\phi}$ is again extremely close to the $tanh$ form, and the width of the domain wall is close to the width of the exact solution. There are also some obvious shortcomings: The simple model does not capture the shifts of the $tanh$ profile for $\hat T_{\tau\tau}$ and $\hat T_{\phi\phi}$ (i.e., a nonzero value of $x_0$ in~\eqref{eq:tanhfit}) which are linked to nonzero surface tension. Moreover, we did not find a way to derive systematically the approximation of equations~\eqref{eq:xiLap} and~\eqref{eq:Jxidef} from the Einstein equations. While the Ansatz of equation~\eqref{eq:modelAn} solves the current conservation equations assuming~\eqref{eq:xiLap}, some of the other Einstein equations are not satisfied.

We will therefore shift our considerations to the boundary theory and proceed to model directly the energy-momentum tensor of the coexisting phases and domain walls.

\section{A covariant energy-momentum tensor for the domain wall and coexisting phases} \label{sec.covdescript}

As explained in the introduction, our goal is to find a simple covariant formulation which would allow us to describe a system made up of coexisting domains of the confining and deconfined phases and domain walls separating them. At the same time we would like to retain the possibility of describing arbitrary plasma flows in the deconfined phase, hence our formulation should necessarily extend hydrodynamics.

Ideally, following the approach of fluid/gravity duality, we would model the holographic spacetime geometry locally as either a patch of the confining geometry or fluid/gravity geometry or a plasma-ball domain wall. Unfortunately, the gravitational description of such a composite system is extremely complicated, as we indicated in the previous section. Therefore, we decided to perform the modelling at the level of the energy-momentum tensor of the boundary field theory, and use the known numerical plasma-ball domain wall solution as additional holographic input which would constrain both the overall structure and the precise coefficients of the field theory energy-momentum tensor.

\subsection{Energy-momentum tensors --- general structure}
\label{s.tmunustructure}

The energy-momentum tensors of the two phases are well known. For the confining phase in the 
 $d=3$ 
Witten's  model~\cite{Witten:1998zw} we have
\eq
\label{e.tmunuconf}
T_{\mu\nu}^{conf} = \eta_{\mu\nu} -4 n_\mu n_\nu
\eqx
where $n^\mu$ is a unit vector pointing in the direction of the auxiliary $\phi$ circle.
As the energy-momentum tensor in the deconfined phase we take the leading order perfect fluid hydrodynamics:
\eq
\label{e.tmunudeconf}
T_{\mu\nu}^{deconf} = p_{hydro}(T) \left( \eta_{\mu\nu} + 4u_\mu u_\nu \right) \ .
\eqx
In this paper we neglect all dissipative terms.
$p_{hydro}(T)$ is the hydrodynamic pressure expressed as a function of the temperature. 
With the normalization of (\ref{e.tmunuconf}), we have
\eq
p_{hydro}(T_c) = 1 \ .
\eqx

Let us now move to the domain-wall solution reviewed in section~\ref{s.domainwall} and recast its energy-momentum tensor in a covariant manner. To this end, let us identify the tensors that we have at our disposal. Apart from the metric $\eta_{\mu\nu}$, we have the flow velocity in the deconfined phase $u^\mu$ (which corresponds to a fluid at rest for the solution from section~\ref{s.domainwall}), the vector $n^\mu$ in the direction of the $\phi$ circle, and a new ingredient -- a vector perpendicular to the domain wall $v^\mu$ (in the later parts of this paper we will take $v^\mu$ to be proportional to a gradient of a scalar field $v_\mu \propto \partial_\mu \gm$).

For the numerical domain wall solution described in the previous section, the energy-momentum tensor turns out to be formed just from \emph{diagonal}\footnote{There could be a possible non-diagonal quartic term $(u\cdot v) \left(u^\mu v^\nu + v^\mu u^\nu -\f{1}{2} (u\cdot v) \eta^{\mu\nu}\right)$. However for the known domain wall solution this term is identically zero, hence we ignore this term in the present paper.} combinations of the above vectors i.e. it is a linear combination of
\eq
\eta_{\mu\nu}, \qq u_\mu u_\nu, \qq v_\mu v_\nu, \qq n_\mu n_\nu 
\eqx
subject to the condition of tracelessness. Hence we have three independent scalar coefficients. 

Clearly the energy-momentum tensors of the confined and deconfined phase also fit the above structure. Therefore we will parametrize the overall energy-momentum tensor as
\eq
\label{e.tmunugen}
T_{\mu\nu}(x) = T_{\mu\nu}^{mix}(x)
+T_{\mu\nu}^\Sg(x)
\eqx
where $T_{\mu\nu}^{mix}(x)$ is a linear combination of the energy-momentum tensors of the two coexisting phases
\eq
T_{\mu\nu}^{mix}(x) = \Gm(x) T_{\mu\nu}^{conf}(x) + (1-\Gm(x)) T_{\mu\nu}^{deconf}(x) \ .
\eqx
The mixing coefficient $\Gm(x)$ will be an additional \emph{non-hydrodynamic} degree of freedom, which we add to the hydrodynamic degrees of freedom $p_{hydro}$ and $u^\mu$.
The goal of this paper is essentially to provide consistent equations of motion and energy-momentum tensor for this additional degree of freedom.
The scalar $\Gm(x)=1$ in the confined phase and $\Gm(x)=0$ in the deconfined phase, at $T=T_c$. 

The second part of the energy-momentum tensor (\ref{e.tmunugen}),
$T_{\mu\nu}^\Sg(x)$ is assumed to vanish when $\Gm(x)$ is a constant equal to either $0$ or $1$. Hence it has support essentially only on the domain wall. We will see later that it is responsible for describing the surface tension of the wall. We can write it in general as
\eq
T_{\mu\nu}^\Sg = \Sg \left(- \eta_{\mu\nu} + A v_\mu v_\nu -B u_\mu u_\nu -C n_\mu n_\nu \right)
\eqx
where $\Sg$, $A$, $B$ and $C$ are coefficients which should be determined.

By specializing to the static planar domain wall solution, we can deduce a further requirement for $T_{\mu\nu}^\Sg$. Conservation of energy-momentum implies that the $T_{xx}$ component is constant. Since this is automatically satisfied by the first two terms
\eq
\Gm(x) + (1-\Gm(x)) \underbrace{p_{hydro}(T_c)}_{=1} = 1 \ ,
\eqx
it follows that $T_{xx}^\Sg$ is a constant. But since we assumed that
$T_{\mu\nu}^\Sg$ vanishes far away from the domain wall, that constant must be equal to zero. 
Therefore we find that $A = 1$ when evaluated on the domain wall solution, so
\eq
\label{e.tmunusigma}
T_{\mu\nu}^\Sg = \Sg \left(- \eta_{\mu\nu} +  v_\mu v_\nu -B u_\mu u_\nu -C n_\mu n_\nu \right) \ .
\eqx
In our setup $T_{\mu\nu}^\Sg$ has to be traceless, so
\eq
B - C = 3 \ .
\eqx

Let us summarize the structure obtained so far. The energy-momentum tensor of a system combining the confined and deconfined phases has the form
\eq
\label{e.tmunutotal}
T_{\mu\nu}(x) = 
\Gm T_{\mu\nu}^{conf}(x) + (1-\Gm) T_{\mu\nu}^{deconf}(x) 
+ \Sg \left(
- \eta_{\mu\nu} +  v_\mu v_\nu -B u_\mu u_\nu + (3-B) n_\mu n_\nu \right)\ .
\eqx
The dynamical variables are: the hydrodynamic pressure $p_{hydro}$ and flow velocity $u^\mu$ in the deconfined phase, the vector $v^\mu$ and $\Gm$, $\Sg$ and $B$ coefficients. 
We will express the latter four quantities in terms of a single scalar field $\gm$.

\subsection{Energy-momentum tensor -- comparison with holography}
\label{s.tmunufits}

We will now go back to the holographic solution for the numerical domain wall geometry described in the previous section, in order to determine $\Gm$, $\Sg$ and $B$. It turns out that two different fits motivated by the analysis of section~\ref{s.phenomenology} %
are possible. We will discuss them now in turn. 
We start with the most natural one in view of the fit results of section~\ref{s.phenomenology}.

\subsubsection*{Option A}

The coefficient $\Gm$ is treated as an independent scalar field $\gm$ and $\Sg$ and $B$ will turn out to be simple functionals of $\gm$.
The  solution of \cite{Aharony:2005bm} is very well reproduced if
\eq
\label{e.GammaA}
\Gm(x) \equiv \gm(x) =\f{1}{2} +  \f{1}{2} \tanh \left(  \f{q_* x}{2} \right)\ .
\eqx
Then looking at the $T_{xx} - T_{yy}$ component of the  energy momentum tensor of \cite{Aharony:2005bm} we may express $\Sg$ as
\eq
\label{e.sigmaA}
\Sg = c\cdot \f{\gm'^2}{\gm (1-\gm)} = c\cdot \f{q_*^2}{4 \cosh^2 \f{q_* x}{2}}\ .
\eqx
Notice that unlike the fit of section~\ref{s.phenomenology}, this description does not include a shift $x^0$. The shift parameter is, in effect, replaced by the coefficient $c$ here. %
This combination has the interpretation of the surface tension density, whose integral over $x$ gives the surface tension (see section~\ref{s.surfacetension}).
Now examining any other component like $T_{tt}$ we find that
\eq
B = 1 + \Gm = 1 + \gm\ .
\eqx
We expect that the expression for $B$ could be different for other theories.

Before we continue, let us comment on the holographic interpretation of the scalar field $\gm$. Consider the region where $\gm$ is very small. Then the energy-momentum tensor is essentially
\eq
T_{\mu\nu} = T_{\mu\nu}^{deconf} + \gm(x) \left(T_{\mu\nu}^{conf} - T_{\mu\nu}^{deconf} \right) + \OO{\gm^2}\ .
\eqx
Holographically, we can understand $\gm$ as the boundary limit of a linearized perturbation of the planar black hole geometry, and thus a quasi-normal mode (QNM). This quasi-normal mode is slightly nonstandard as 
its behaviour is like
\eq
\label{e.qmnasympt}
q(x) \sim e^{q_* x}
\eqx
so it has purely imaginary momentum and vanishing frequency.
Indeed such QNM's were identified in the context of nonequilibrium steady states in~\cite{Sonner}, with the imaginary momentum setting the scale for the characteristic width, and earlier in the context of an absorption length in~\cite{Amado:2007pv}. As this holographic degree of freedom is crucial for the description of a domain wall, and thus we should incorporate it directly into any low-energy effective description. 
We give some additional comments on the possible interpretation of $\gm$ in section~\ref{s.generality}.

\subsubsection*{Option B}

The second option is slightly less natural. Its advantage is its relation to a canonical scalar field action as we will see in the following section as well as a link to Landau's description of phase transitions which we touch on in section~\ref{s.generality}. 

Since as we saw in section~\ref{s.phenomenology}, the $T_{xx} - T_{yy}$ can also be fit by $1/\cosh^4$, which allows for taking
\eq
\Sg = c \cdot \gm'^2 = c \cdot \f{q_*^2}{16 \cosh^4 \f{q_* x}{2}}
\eqx
without the $\gm(1-\gm)$ denominator appearing in (\ref{e.sigmaA}). 
The drawback is that the $q_*$ coming from the above expression is not compatible with the mixing coefficient $\Gm=\gm$ and (\ref{e.GammaA}). One can nevertheless cure this discrepancy by taking the following more involved expression for~$\Gm$:
\eq \label{e.GammaB}
\Gm = \gm^2 (3-2\gm)\ .
\eqx
This explicit form is motivated by the considerations of section~\ref{s.acc}.
The coefficient $B$ is again given by
\eq
B = 1 + \Gm\ .
\eqx

\begin{figure}[t!]
\begin{center}
\includegraphics[width=.8\textwidth]{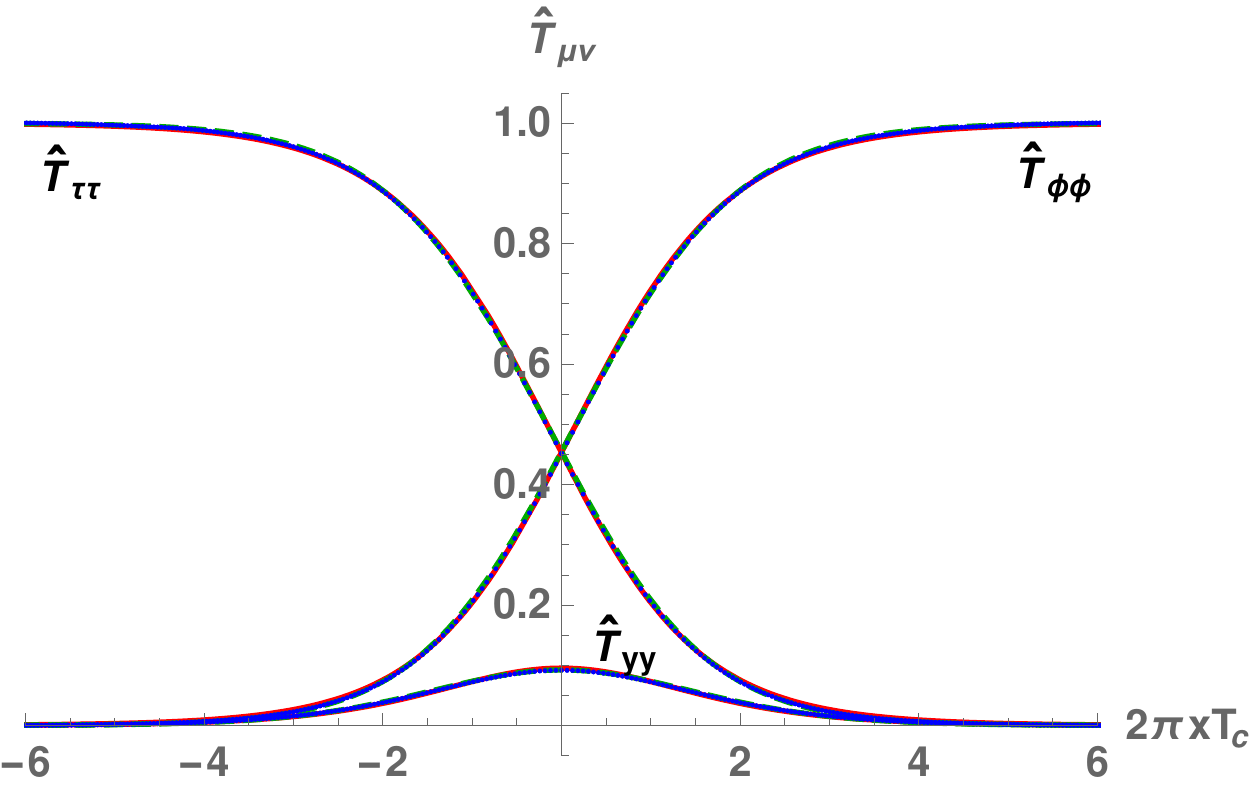}
\end{center}
\caption{Final fits of the model to the numerical solution for the domain wall. The blue dots, solid red curve, and dashed green curve show the numerical data, fit using the option A, and fit using the option B, respectively. \label{fig:Tmunufinal}}
\end{figure}

\subsubsection*{Comparison with the numerical solution}

We then compare our formulation to the components of the energy momentum tensor extracted from the numerical solution for the domain wall~\cite{Aharony:2005bm} which we discussed in section~\ref{s.domainwall_numerical}. By using the expression for the complete energy momentum tensor in~\eqref{e.tmunutotal} (and transforming to Euclidean conventions), we obtain for the normalized components $\hat T_{\mu\nu}$ defined in~\eqref{eq:hatTmunudef}
\be \label{eq:TmunuGammaSigma}
 \hat T_{\tau\tau} = 1-\Gamma - \frac{1}{4}\Sigma \Gamma \ , \qquad 
  \hat T_{\phi\phi} = \Gamma - \frac{1}{4}\Sigma(1- \Gamma) \ , \qquad 
   \hat T_{yy} = \frac{1}{4}\Sigma \ ,
\ee
where we already inserted $B=1+\Gamma$. We remind that for the two options discussed above we have
\begin{align}
\label{e.optionA}
    \Gamma &= \gamma \ , &  \Sigma &= c\, \frac{\gamma'^2}{\gamma(1-\gamma)} \ ,& &\textrm{(Option A)} \ ;& \\
    \label{e.optionB}
    \Gamma &= \gamma^2(3-2\gamma) \ , &  \Sigma &= c\, \gamma'^2 \ ,& &\textrm{(Option B)} \ , & 
\end{align}
and for both options 
\be 
\label{e.gammaoptions}
\gamma(x) = \f{1}{2} +  \f{1}{2} \tanh \left(  \f{q_* x}{2} \right) \ . \ee

We have carried out a least squared fit of $q_*$ and $c$ to the data for $\hat T_{\mu\nu}$ extracted from the numerical domain wall solution to the Einstein equations. We used the components $\hat T_{\tau\tau}$ and $\hat T_{yy}$ for the fit (and excluded $\hat T_{\phi\phi}$ because it gives no additional information due to reflection symmetry about the center of the wall). The fit results are shown in figure~\ref{fig:Tmunufinal}. The fit for option~B is slightly better than the fit for option~A, with 30\% smaller root mean square error. It is however difficult to see this from the plot because both curves are so close to the data. Actually, the deviation from the numerical data is smaller than its accuracy for both fits.

The fit results are given by
\begin{align}
    \frac{q_*}{2\pi T_c} & \approx 1.055 \ , \qquad c\, (2\pi T_c)^2 \approx 0.342 \ , & &\textrm{(Option A)} \ ;& \\
    \frac{q_*}{2\pi T_c} & \approx 0.682 \ , \qquad c\, (2\pi T_c)^2 \approx 3.167 \ , & &\textrm{(Option B)} \ .& 
\end{align}
We note that the value of $q_*$ for option~A (option~B) agrees well with the fit result to $\hat T_{yy}$ carried out in section~\ref{s.phenomenology} with the $cosh^{-2}$ ($cosh^{-4}$) Ansatz, i.e., $q_*^{(2)}$ ($q_*^{(4)}$) in equation~\eqref{eq:cfitresd3}. This is as expected: using the $tanh$ form for $\gamma$ leads to $\hat T_{yy} \propto \cosh^{-2}(q_*x/2)$ for option~A and   to $\hat T_{yy} \propto \cosh^{-4}(q_*x/2)$ for option~B. Therefore the main difference between the final fit and the fits of section~\ref{s.phenomenology} is that the final fit is able to describe all components of $T_{\mu\nu}$ in terms of a single value of $q_*$. For option~A, the (relatively small) difference between the values of $q_*$ for the separate fits of the components in section~\ref{s.phenomenology} is alleviated here by a smart choice of the ``subleading'' $\propto \Sigma$ terms for $\hat T_{\tau\tau}$ and $\hat T_{\phi\phi}$ in~\eqref{eq:TmunuGammaSigma}, which arises in turn from the choice of the function $B(\Gamma)$. For option~B, the (much larger) difference between the values of $q_*$ was removed by an appropriate choice of $\Gamma(\gamma)$ as we already pointed out above.  

\subsubsection*{Equation of motion for $\gm$}

The above fitted expressions relied on the assumption that for the  domain wall solution of~\cite{Aharony:2005bm}, $\gm$ is given by a $tanh$ function:
\eq
\label{e.gmtanh}
\gm(x) =\f{1}{2} + \f{1}{2} \tanh \left( \f{q_* x}{2} \right)
\eqx
Of course, this should not be an assumption but should follow from some specific equations of motion. It's well known that (\ref{e.gmtanh}) is a solution of the virial theorem
\eq
\label{e.virialorg}
\gm' = \sqrt{2 V(\gm)} \qq \text{with} \qq V(\gm) = \f{q_*^2}{2} \gm^2 (1-\gm)^2\ .
\eqx
Indeed (\ref{e.virialorg}) is a solution of the equations of motion for any scalar action of the form
\eq
\label{e.preliminary}
\LL_\gm^{preliminary} = - a(\gm) \left(  \f{1}{2} (\partial \gm)^2 +V(\gm) \right)
\eqx
with arbitrary prefactor function $a(\gm)$.
The Lagrangian (\ref{e.preliminary}) cannot be the whole story, however, as the energy-momentum tensor derived from (\ref{e.preliminary}) does not include the $u^\mu u^\nu$ terms appearing in (\ref{e.tmunutotal}).
We thus need to couple the scalar action to hydrodynamics in the deconfined phase.
We will perform this construction in the following section and provide an action which will generate equations of motion for $\gm$ and reproduce the structure of the energy-momentum tensor\footnote{Under the assumption of $\gm(x)$ satisfying the virial theorem.} (\ref{e.tmunutotal}). Together with the standard energy-momentum conservation equations for the hydrodynamic variables we will thus have a fully specified system.

\section{An action density and the final formulation}
\label{s.action}

In this section we will 
propose a general from of an action for $\gm(x)$, whose corresponding energy-momentum tensor will be capable of reproducing the structure (\ref{e.tmunusigma}).

Before we do that, we will review an action formulation for perfect fluid hydrodynamics as it will provide the natural ingredients for introducing the dependence on hydrodynamics into the action for $\gm(x)$.

\subsection{An action for perfect fluid hydrodynamics}

An action for perfect fluid hydrodynamics was first introduced in \cite{Hydroaction}, we will, however, use a later formulation by \cite{Mukund} as it matches with the on-shell action in holography.

Following this formulation, as the elementary hydrodynamic degree of freedom one takes the vector $\beta^\mu$ related to the temperature through
\eq
T=\f{1}{\sqrt{-g_{\mu\nu} \bt^\mu \bt^\nu}}\ .
\eqx
We get then the following variation, expressed in terms of the conventional flow velocity $u^\mu = T \bt^\mu$:
\eq
\dl T = -\f{1}{2} T u_\mu u_\nu \dl g^{\mu\nu}\ .
\eqx
The hydrodynamic Lagrangian is then simply given by the pressure expressed as a function of the temperature
\eq
\LL_{hydro} = p(T)\ .
\eqx
Using the formula
\eq
T_{\mu\nu} = \f{-2}{\sqrt{-g}} \f{\dl(\sqrt{-g} \LL)}{\dl g^{\mu\nu}} = -2 \f{\dl \LL}{\dl g^{\mu\nu}} + g_{\mu\nu} \LL
\eqx
one recovers the hydrodynamic perfect fluid energy-momentum tensor
\eq
T_{\mu\nu}^{hydro} = \underbrace{T \partial_T p}_{\eps+p}  u_\mu u_\nu + p g_{\mu\nu}\ .
\eqx
The linear combination of the hydrodynamic and confining energy-momentum tensors would then follow from the Lagrangian
\eq
\label{e.hydroconf}
\LL = (1-\Gm) p(T) + \Gm\ .
\eqx

\subsection{The action for the scalar field $\gm$} \label{sec:gammaaction}

We will primarily consider a formulation just in terms of the \emph{physical} 3D coordinates, ignoring the auxiliary angular $\phi$ coordinate as we assume no dependence on $\phi$. For this reason, we will not recover now the $n^\mu n^\nu$ terms in the energy-momentum tensor. We will show in Appendix~\ref{s.nmunnu}, how to formally reproduce also those terms. 

Consider a rather general scalar field action coupled to the hydrodynamic degrees of freedom through a dependence on $T$
\eq
\LL_\gm = -\f{1}{2} a(\gm, T) g^{\mu\nu} \partial_\mu \gm \partial_\nu \gm - b(\gm, T)\ .
\eqx
We ignore possible terms with $\bt^\mu \partial_\mu \gm$ as we do not have any information about them from known gravity solutions.
The resulting energy momentum tensor is
\eq
\label{e.tmunuscalar}
T_{\mu\nu}^\Sg = a \partial_\mu\gm \partial_\nu\gm -\left( \f{1}{2}a (\partial \gm)^2 +b \right) g_{\mu\nu}
- T \left( (\partial \gm)^2 \partial_T a  + \partial_T b \right) u_\mu u_\nu\ .
\eqx
Let us now specialize to the domain wall solution from section~\ref{s.phenomenology} and impose the constraint
\eq
\label{e.txxsg}
T_{xx}^\Sg = 0\ .
\eqx
Recall from the discussion in section \ref{s.tmunufits} that
for the planar domain wall solution at $T=T_c$ we would like to have 
\eq
\label{e.virial}
\gm' =\sqrt{2 V(\gm)}\ .
\eqx
Plugging this into (\ref{e.txxsg}) gives
\eq
a\gm'^2 -\f{1}{2}a \gm'^2 -b =0
\eqx
which yields (at $T=T_c$)
\eq
\label{e.b}
b = a V\ .
\eqx
So the scalar action becomes (again the argument holds strictly speaking just at $T=T_c$)
\eq \label{eq:gammaaction}
\LL_\gm = - a(\gm, T) \left(  \f{1}{2} (\partial \gm)^2 +V(\gm, T) \right)\ .
\eqx
Recall 
that for \emph{any} $a(\gm, T)$ the virial theorem solution (\ref{e.virial}) solves the equations of motion coming from this action.

Let us now, for simplicity, assume an overall power-law dependence on the temperature\footnote{Of course this should be treated just as an approximation close to $T_c$ and not as a fundamental assumption.}, and take $T_c=1$, i.e.
\eq
\label{e.aV}
a(\gm, T) = T^\alpha a(\gm)\ , \qqqq
V(\gm, T) = T^\beta V(\gm)\ .
\eqx
Evaluating (\ref{e.tmunuscalar}) using the virial theorem (\ref{e.virial}) at $T=1$ we get
\eq
T_{\mu\nu}^\Sg = a \left[ \underbrace{\partial_\mu\gm \partial_\nu\gm}_{(\partial \gm)^2 v_\mu v_\nu}  - (\partial \gm)^2 \eta_{\mu\nu} - (\partial \gm)^2 \left(  \f{3}{2} \alpha + \f{1}{2} \beta   \right) u_\mu u_\nu \right]\ .
\eqx
We see that we recover the structure of (\ref{e.tmunusigma}), when specializing to a solution satisfying the virial theorem (\ref{e.virial}). Otherwise we have the general form (\ref{e.tmunuscalar}) with (\ref{e.b}) and (\ref{e.aV}).
For a formal extension of the above treatment which includes the $n^\mu n^\nu$ terms in the energy-momentum tensor see Appendix~\ref{s.nmunnu}.
The coefficient $B$ appearing in (\ref{e.tmunusigma}) is expressed as
\eq
\label{e.Balphabeta}
B =  \f{3}{2} \alpha + \f{1}{2} \beta \ .
\eqx
From the planar domain wall solution that we have at our disposal, we cannot fix the individual coefficients $\alpha$ and $\beta$ apart from the above linear combination. We leave their determination for future study. 
We will however present an argument in section~\ref{s.acc} which suggests that the possible $\gamma$ dependence of $B$ arises solely through the coefficient $\alpha$.
The coefficient $\beta$ essentially characterises how the width of the domain wall changes as we move away from $T=T_c$.
The surface tension density is given by
\eq
\Sg = a(\gm) \left( \f{1}{2} (\partial \gm)^2 +V(\gm) \right) \longrightarrow a(\gm) (\partial \gm)^2
\eqx
where the latter form follows when using the virial theorem.

\subsection{The final formulation}
\label{s.final}

Let us summarize here our final form of the action
\eq
\label{e.actionfinal}
\LL_{final} = (1-\Gm(\gm)) p(T) + \Gm(\gm) 
- a(\gm, T) \left(  \f{1}{2} (\partial \gm)^2 +V(\gm, T) \right)
\eqx
with $a(\gm, T) = T^\alpha a(\gm)$, and $V(\gm, T) = T^\beta V(\gm)$ understood as approximate expressions for temperatures close to $T_c=1$.

In order to fully specify the above action, we need to pick the prefactor $a(\gm)$ and the potential $V(\gm)$. Furthermore, we need to express $\Gm(\gm)$ in terms of the elementary scalar field~$\gm$. The fits of the energy-momentum tensor presented in section~\ref{s.tmunufits} yield the two natural options
\begin{align}
\label{e.optionAfinal}
    \Gamma(\gm) &= \gamma \ , &  a(\gm) &= c \frac{1}{\gamma(1-\gamma)} \ ,& &\textrm{(Option A)} \ ;& \\
\label{e.optionBfinal}    
\Gamma(\gm) &= \gamma^2(3-2\gamma) \ , &  a(\gm) &= c  \ ,& &\textrm{(Option B)} \ , & 
\end{align}
with the potential
\eq
\label{e.potfinal}
V(\gm) = \f{q_*^2}{2} \gm^2 (1-\gm)^2\ .
\eqx
The $\al$ and $\bt$ coefficients characterize how the parameters of the scalar action behave as we move away from $T=T_c$. Since we have numerical data purely at $T=T_c$, we can currently fix only their linear combination (\ref{e.Balphabeta}) which is given by the function $B(\Gm)$. From a numerical fit for the $d=3$ Witten model we have
\eq
B(\Gm) = 1 + \Gm\ .
\eqx
For other models we expect that the expression for $B(\Gm)$ will get modified (see Appendix~\ref{app:d4} and \ref{s.nonconformal} for explicit examples). On the other hand, even for those different models, the expressions (\ref{e.optionAfinal})-(\ref{e.potfinal}) can remain unchanged.

Finally let us note that even though the potential appearing directly in the scalar action~(\ref{e.potfinal}) remains symmetric under the interchange $\gm \leftrightarrow 1-\gm$ even as we move away from $T=T_c$, the overall potential for $\gm$ automatically breaks this symmetry (as it should!) due to the first two terms in (\ref{e.actionfinal}) describing the individual phases of the theory.   
We will return to this point later in section~\ref{s.coupling} and in section~\ref{s.generality} from a more general perspective.

\section{Some generic physical applications}

We will now illustrate the general structure of the proposed energy-momentum tensor (\ref{e.tmunutotal}) with a couple of applications.
Here we focus on considerations which are very generic and depend just on the form of the energy-momentum tensor and not on the details of the formulation proposed in section~\ref{s.final}.

Firstly, we link the parametrization of the energy-momentum tensor with the surface tension of the domain wall. Then we compare equations for an equilibrium circular droplet with standard thermodynamic considerations.
Finally, we reproduce the formula for thermodynamic nucleation probability from \cite{Landau} using Euclidean solutions as suggested by Linde~\cite{Linde}.

In this section it is convenient to use the notation
\eq
P(x) \equiv \Gm(x) + (1-\Gm(x)) p_{hydro}(T)
\eqx
for the coefficient of $\eta_{\mu\nu}$ in $T_{\mu\nu}^{mix}(x)$:
\eqn
\label{e.tmunuP}
T_{\mu\nu}^{mix}(x) &=& \left(\Gm(x) + (1-\Gm(x)) p_{hydro}(T) \right) \eta_{\mu\nu} + 4(1-\Gm(x)) p_{hydro}(T) u_\mu u_\nu -4 \Gm(x) n_\mu n_\nu \nonumber\\
& = & P(x) \eta_{\mu\nu} + 4\left(P(x) - \Gm(x)\right) u_\mu u_\nu -4\Gm(x) n_\mu n_\nu\ .
\eqnx
$P(x)$ can be interpreted as an overall pressure valid in both phases.

Since in the $d=3$ Witten model the physical $d=3$ theory is dimensionally reduced on the~$\phi$ circle from the auxiliary boundary $d=4$ theory, one has to take care for which theory the particular energy-momentum tensor is defined.
Up to this point, we always considered the energy momentum tensor for the full auxiliary $d=4$ theory, which would be the expression read off directly from the holographic dual geometry. In this section we will focus on applications dealing with the physical dimensionally reduced theory, hence we should dimensionally reduce the energy-momentum tensor as well.

In order to avoid cumbersome notation, here we will use units such that the range of the~$\phi$ circle coordinate is equal to 1. Then the components of the physical energy momentum tensor will be 
\eq
\label{e.phi3d}
T_{\mu\nu}^{3d} (x) = \int T_{\mu\nu}^{4d} (x)\, d\phi = \left( \int d\phi \right) \cdot T_{\mu\nu}^{4d} (x)
\eqx
so $T_{\mu\nu}^{3d}$ is just obtained from $T_{\mu\nu}^{4d}$ by dropping the $n_\mu n_\nu$ term.
Hence we do not need to explicitly add the ${}^{3d}$ or ${}^{4d}$ superscripts.

\subsection{The domain wall surface tension}
\label{s.surfacetension}

The surface tension extracted from the planar domain wall solution is given by the following integral
\eq
\SSg = \int_{-\infty}^\infty T_{xx} - T_{yy} \;dx
\eqx
where $y$ is the direction along the domain wall, while $x$ is the perpendicular direction. Let us comment on the units. The elements of the energy momentum tensor for the auxiliary boundary theory have dimension $energy^4$. We integrate over $\phi$ as in (\ref{e.phi3d}) and then over $x$ so $\SSg$ has dimension $energy^2$ which is the dimension of the surface tension in a 3-dimensional theory. Plugging in (\ref{e.tmunutotal}) we get
\eq
\SSg = \int_{-\infty}^\infty \Sg(x) \; dx
\eqx
which motivates the notation for $\Sg(x)$. Of course,
following our assumptions on the support of $T_{\mu\nu}^\Sg(x)$
this integral gets contributions essentially just from the region of the domain wall. This indeed holds in the concrete models proposed in section~\ref{s.final}.

Let us make a further comment here. We saw that the $T_{\mu\nu}^\Sg$ could arise as the energy-momentum tensor coming from an effective scalar field theory through
\eq
\label{e.tmunudef}
T_{\mu\nu} = \f{-2}{\sqrt{-g}} \f{\dl(\sqrt{-g} \LL)}{\dl g^{\mu\nu}} = -2 \f{\dl \LL}{\dl g^{\mu\nu}} + g_{\mu\nu} \LL \ .
\eqx
Hence the coefficient of $\eta_{\mu\nu}$ in $T_{\mu\nu}^\Sg$ (recall (\ref{e.tmunusigma})) has the interpretation of the on-shell action density evaluated on the domain wall solution.
As for static solutions the Euclidean action is minus the Minkowski one, the surface tension would be identified with the value of the Euclidean scalar field action evaluated on the domain wall solution. This is indeed a very natural behaviour.

\subsection{A circular droplet}

Let us first recall the standard thermodynamic analysis of a droplet of one phase in another phase \cite{Landau}
adapted to the current dimensionality ($d=2+1$).
Suppose that we have a droplet of radius $R$ of the phase with pressure $P_{droplet}$. The spatial volume (here area) of the droplet is $V_{droplet}=\pi R^2$,
and its circumference is $A=2\pi R$. The surface tension of the interface is~$\SSg$. The environment has volume $V$ and pressure $P$.
The total thermodynamic potential of the system is given by
\eq
\Omega = -P_{droplet} V_{droplet} -P V + \SSg A \ .
\eqx
We should extremize the above formula w.r.t. $R$, keeping in mind that $dV/dR = -dV_{droplet}/dR$ as $V+V_{droplet}=V_{total}=const$. We get
\eq
\f{\partial \Omega}{\partial R} = -(P_{droplet}-P) 2 \pi R + 2 \pi \SSg = 0 \ .
\eqx
Hence we obtain the equation for the equilibrium size of the droplet
\eq
\label{e.thermodynamicR}
P_{droplet} - P = \f{\SSg}{R}\ .
\eqx 
We will now proceed to show how the above equation arises from (\ref{e.tmunutotal}) in the thin wall approximation.

We will use here the polar coordinate system as in the present case the spherical domain wall is 
really circular since we have just two spatial coordinates for the physical theory.
\eq
ds^2 = -dt^2 + dr^2 +r^2 d\th^2 + d\phi^2\ .
\eqx
Recall that $\phi$ is the auxiliary coordinate of the Witten's circle and we assume that nothing depends on it. The physical coordinates of the theory are $t$, $r$ and $\th$.
The nontrivial conservation law takes the form
\eq
\partial_r T^{rr} -r T^{\th\th} +\f{1}{r}T^{rr} = 0\ .
\eqx
The vector fields are $u^\mu=(1,0,0,0)$ and $v^\mu=(0,1,0,0)$ and the relevant components of the energy-momentum tensor~\eqref{e.tmunutotal} are
\eq
T^{\th\th} = \f{P - \Sg}{r^2} \ ,  \qqqq T^{rr} = P -\Sg +\Sg = P\ .
\eqx
We thus get
\eq
\label{e.circeq}
\partial_r P = -\f{1}{r} \Sg\ .
\eqx
In the thin-wall approximation we assume that the size of the droplet is much larger than the width of the domain wall. Since $\Sg$ is nonzero only in the vicinity of the domain wall, we may replace in this approximation $1/r$ by $1/R$. Then integrating both sides of equation (\ref{e.circeq}) we obtain
\eq
\underbrace{P(r=+\infty) - P(r=0)}_{P - P_{droplet}} = -\f{\SSg}{R} 
\eqx
which reproduces (\ref{e.thermodynamicR}).

Of course, in order to compute the value of $\SSg$ and obtain the full profile $P(r)$ and not only the asymptotic values, we need to specify the expression of $\Sg$ in terms of a scalar field $\gm$ and solve the equations of motion for $\gm$ following from (\ref{e.actionfinal}).
However for reproducing the thermodynamic equilibrium condition (\ref{e.thermodynamicR}), these specific details are not necessary.

\subsection{Thermodynamic nucleation probability and Euclidean solutions}

The nucleation probability of a bubble of one phase in an environment of the other phase can be computed in two ways.
On the one hand, one can use standard thermodynamic arguments of \cite{Landau} which make use of the value of the surface tension.
On the other hand, one can use the field theoretical perspective of Linde \cite{Linde}, which is a finite temperature generalization of the considerations of Coleman~\cite{Coleman:1977py, Callan:1977pt}.
In this approach one constructs a solution of the Euclidean equations of motion for a configuration of a static bubble of one phase embedded in the other phase with standard Euclidean time periodicity. 

Finding the relevant Euclidean solution within holography is a formidable task which would require using nontrivial numerical relativity methods, especially in the context of confining and deconfined phases described within the Witten's model~\cite{Witten:1998zw} (or use metric Ans\"atze like in~\cite{Bigazzi:2020phm}).
However, since our boundary formulation in terms of the energy-momentum tensor is in essence a shortcut to the dual holographic background\footnote{In the same sense as from hydrodynamics one can get to the dual gravitational background through fluid/gravity duality.}, we can just solve instead equations on the boundary. Moreover, it turns out that to get the nucleation probability in the thin wall approximation, we do not need to find any explicit solutions at all.

Let us first recall the thermodynamic approach from \cite{Landau}. The probability 
of a thermal fluctuation producing a nucleus of another phase
is given by
\eq
\label{e.probrmin}
probability \propto e^{-R_{min}/T}
\eqx
where $R_{min}$ is the minimal work to form the nucleus (this follows from the probability of a fluctuation of a closed system $\propto e^{S}$).
This can be expressed in terms of the thermodynamic potentials of the uniform phase
\eq
\label{e.before}
\Omega_{before} = -P (V + V_{droplet})
\eqx
and of the system with the droplet
\eq
\label{e.after}
\Omega_{after} = -PV -P_{droplet}V_{droplet} + \SSg A\ .
\eqx
Then 
\eq
\label{e.rmin}
R_{min} = \Omega_{after} - \Omega_{before} = -(P_{droplet} - P) V_{droplet} + \SSg A
\eqx
so
\eq
\label{e.probability}
probability \propto e^{-\f{1}{T} \left( -(P_{droplet} - P) V_{droplet} + \SSg A \right)} \ .
\eqx
This formula should be applied to the \emph{equilibrium} droplet with size $R$ given by (\ref{e.thermodynamicR}).

Let us now show how one can reproduce (\ref{e.probability}) from Linde's approach employing our framework. In principle we should find a gravitational solution for an equilibrium droplet of size $R$, compute its Euclidean on-shell gravitational action and subtract the action of the uniform phase without a droplet. All this quite nontrivial.
As explained earlier, we can sidestep these difficulties using the proposed general structure of the energy-momentum tensor. 

Clearly, the key ingredients to reproduce are the thermodynamic potentials (\ref{e.before}) and (\ref{e.after}), which, when divided by $T$ should coincide with the values of the \emph{Euclidean} on-shell action for the field configurations.

\emph{A-priori} the knowledge of just the energy-momentum tensor does not allow to obtain the on-shell action. However if we assume, as is the case in our construction, that the energy-momentum tensor arises from some Lagrangian density through (\ref{e.tmunudef}), we see that we can read off the Minkowski Lagrangian density from the coefficient of $\eta_{\mu\nu}$, change sign when passing to Euclidean signature and integrate it over Euclidean spacetime to obtain the on-shell thermal action.

Consequently, from (\ref{e.tmunuP}) and (\ref{e.tmunusigma}) we get for the on-shell value of the Euclidean Lagrangian density
\eq
\LL_E(x) = -P(x) + \Sg(x)\ .
\eqx
Hence the on-shell action for the uniform phase would be
\eq
\label{e.Sbefore}
S_{before}^{on-shell} = \int \LL_E(x) d\tau d^2x d\phi = - \f{1}{T} P (V + V_{droplet})
\eqx
where $V + V_{droplet}$ is just the total volume and $\Sg=0$ since we are in the uniform phase.
The on-shell action for the circular droplet configuration in the thin wall approximation (i.e. at large $R$) would then be
\eq
\label{e.Safter}
S_{after}^{on-shell} = \int \LL_E(x) d\tau d^2x d\phi = - \f{1}{T} \left(P V +P_{droplet} V_{droplet} - \SSg A  \right)
\eqx
where the integral of $\Sg(x)$ would get contribution only from the domain wall and in the large $R$ limit integrate to the surface tension times the bubble boundary area. We neglect corrections from the variation of $P(x)$ within the domain wall region as they would be suppressed by $1/R$. 
Taking the difference of (\ref{e.Safter}) and (\ref{e.Sbefore}), we reproduce the thermodynamic nucleation probability (\ref{e.probability}).

Again we see that in the thin wall approximation of large bubble size, we do not need to specify any microscopic details and just the overall knowledge of the surface tension suffices.
We can view the above computations as a physical consistency check of the proposed general structure of the energy-momentum tensor (\ref{e.tmunutotal})  and the interpretation of the coefficient of $\eta_{\mu\nu}$ as the on-shell action density.

\section{Accelerating planar domain wall solution}
\label{s.acc}

As we move away from $T=T_c$, the pressures of the two phases on both sides of the domain wall will no longer balance, and the domain wall will start to move with a uniform acceleration. It is instructive to consider this case, as it gives a nontrivial application (and consistency check) of the proposed framework but also leads to some further constraints on $\Gm(\gm)$ when $a(\gm)=1$.
As it turns out, such accelerating solutions may be obtained simply by moving away from $T = T_c$ in the hydrodynamic part of the action. Namely, the asymmetry imposed by the coupling of the $\gamma$ field to the hydrodynamic action provides exactly the necessary asymmetry in the effective potential for $\gamma$.

\subsection{Effective actions and accelerating solutions}\label{sec:effacc}

However, before considering the coupling to hydrodynamics, let us first recall how accelerating domain wall solutions arise from simple effective actions. We consider small $T -T_c \propto \eps$. Assuming a standard kinetic term (as in option~B above), i.e., setting $a(\gamma,T)=1$ in the scalar action~\eqref{eq:gammaaction}, we search for a generalization of the potential~\eqref{e.potfinal}. If we require that the locations of the two minima remain at $\gamma=0$ and $\gamma=1$, the unique fourth order polynomial (up to $\mathcal{O}(\eps)$ corrections in the normalization and a trivial constant term) which works is
\be
 V_\eps(\gamma) =  \frac{q_*^2}{2}(\gamma-1)^2\left[\gamma^2-\frac{\eps}{3q_*}(1+2\gamma)\right] \ .
\label{eq:Vepsdef}
\ee
This definition also determines the normalization of $\eps$.

We may then search for accelerating solutions by adding $\mathcal{O}(\eps)$ correction terms to the static domain wall solution. Interestingly, a very simple solution exists, and the perturbed solution may be conveniently written as
\be
  \gamma(t,x) =  \frac{1}{2}\left[1+\tanh\left(q_*\left(\frac{x-x_0}{2}-\frac{\eps\, t^2}{4}\right)\right)\right] \ . \label{eq:gammatsol}
\ee
This solution is however only valid for small $\eps \,t^2$, otherwise the perturbation grows large. Knowing the solution near $t=0$ is however enough for the full time evolution because we can always use Lorent covariance to boost to a frame where the domain wall is almost at rest. That is, it is enough to write the result in an explicitly covariant form to capture the full time evolution at $\mathcal{O}(\eps)$. 
A covariant way to write the result is 
\be \label{eq:gammatcov}
  \gamma(t,x) =   \frac{1}{2}\bigg[1  +\tanh\Big(\frac{q_*}{2}\left(\sqrt{(x-x_0-\eps^{-1})^2-t^2}-\eps^{-1}\right)\Big)\bigg] \ .
\ee
For this expression the convergence as $\eps \to 0$ is uniform over all $x$ and $t$.

Above we pointed out that the static $tanh$ solution for the domain wall is obtained for the scalar action of~\eqref{eq:gammaaction} for any choice of $a(\gamma,T)$ if the potential is the standard choice of~\eqref{e.potfinal}. Therefore one might expect that accelerating solution of~\eqref{eq:gammatcov} also generalizes to nontrivial choices of $a(\gamma,T)$. This turns out indeed to be the case. The potential however needs to be different from~\eqref{eq:Vepsdef} if we want the solution to remain in simple form. 

We  consider here only the case $a(\gamma,T) = c/\gamma/(1-\gamma)$, corresponding to option~A above. For this choice, one can indeed easily check that the potential of~\eqref{eq:Vepsdef} does not work. Instead, the potential in this case must have zeroes at $\gamma=0$ and $\gamma=1$ that cancel the poles of $a$ such that the product $a V$ is regular. Therefore we write 
\be \label{eq:VepsA}
 V_\eps(\gamma) = \frac{q_*^2}{2}\gamma(1-\gamma)^2\left(\gamma-\frac{2}{q_*} \eps\right) \ .
\ee

Since the potential term appearing in the full equation of motion is
\be
 \propto \frac{1}{a(\gamma)} \frac{d}{d\gamma}\left[a(\gamma)V(\gamma)\right] \propto \gamma(1-\gamma) \label{eq:potterms}
\ee
we expect that the asymptotic values of the domain wall solution are exactly $\gamma = 0$ and $\gamma = 1$ even after adding the perturbation. And indeed we may verify that~\eqref{eq:gammatcov} is a solution for this potential up to terms $\mathcal{O}(\eps)$. In particular, we chose the normalization of the perturbation term in~\eqref{eq:VepsA} such that it the acceleration matches with that given by constant $a$ and potential~\eqref{eq:Vepsdef}.

\subsection{Coupling to hydrodynamics}
\label{s.coupling}

We then consider the coupling between $\gamma$ and the hydrodynamic degrees of freedom. As mentioned above, these couplings will in effect lead to asymmetric terms for the potential of $\gamma$, eliminating the need of introducing such terms explicitly. 

We use the full hydrodynamic action, given in~\eqref{e.actionfinal} above, with the definitions in~\eqref{e.optionAfinal}--\eqref{e.potfinal}. 
The hydrodynamic terms in the action can be rearranged as
\begin{align}
 (1-\Gamma)p(T)+ \Gamma &= p(T) - (p(T)-1)\Gamma = p(T) - \Delta p(T)\Gamma  
\end{align}
where   
$\Delta p(T) = p(T)-1$ is the pressure difference between the deconfined and confined phases.
The $\Gamma$ term in the action may be interpreted as a change in the potential for $\gamma$:
\be
 a(\gamma)V(\gamma) \to a(\gamma)V(\gamma) + \Delta p(T)(\Gamma-1) \equiv a(\gamma)V_\mathrm{eff}(\gamma) \ .
 \label{eq:aVvar}
\ee 

From here on we need to consider the two options separately. For option~A (i.e., $a(\gamma)=c/\gamma/(1-\gamma)$ and $\Gamma(\gamma)=\gamma$) we find that
\be
 V_\mathrm{eff}(\gamma) = \frac{q_*^2}{2}\gamma^2(1-\gamma)^2 - \frac{\Delta p(T)}{c}\gamma(1-\gamma)^2 =\frac{q_*^2}{2}\gamma(1-\gamma)^2\left(\gamma -\frac{2 \Delta p(T)}{c q_*^2}\right) \ .
\ee
This result, somewhat miraculously, has exactly the same form as~\eqref{eq:VepsA}. Notice that we did not assume that $T-T_c$ is small so far. In the limit of small temperature difference $T-T_c =\mathcal{O}(\eps)$, we find that 
\be
 \Delta p(T) = s_c (T-T_c) +\mathcal{O}(\eps^2) \ ,
\ee
where $s_c=s(T_c)=p'(T_c)$ is the entropy at the critical temperature. 
Comparing to~\eqref{eq:VepsA}, we identify
\be\label{eq:epsoptA}
 \eps = \frac{s_c(T-T_c)}{c q_*} \ . 
\ee
For option~B, i.e., with $a(\gamma)=c$ 
we find that
\be \label{e.VeffB}
 V_\mathrm{eff}(\gamma) = \frac{q_*^2}{2}\gamma^2(1-\gamma)^2 - \frac{\Delta p(T)}{c}(1-\Gamma(\gamma)) \ . 
\ee
For this expression to agree with~\eqref{eq:Vepsdef}, we must have $1-\Gamma(\gamma) \propto (\gamma-1)^2(1+2\gamma)$. This holds if we choose
\be \label{e.GammaoptB}
 \Gamma(\gamma) = \gamma^2(3-2\gamma) \ ,
\ee 
and this expression also satisfied the properties $\Gamma(0)=0$, $\Gamma(1)=1$, and $\Gamma(\gamma)+\Gamma(1-\gamma)=1$ expected from symmetry. We have therefore derived the form introduced for option~B in~\eqref{e.GammaB} above. Inserting it in~\eqref{e.VeffB} we obtain
\be
 V_\mathrm{eff}(\gamma) =\frac{q_*^2}{2}(1-\gamma)^2\left(\gamma^2 -\frac{2 \Delta p(T)}{c q_*^2}(1+2\gamma)\right) \ .
\ee
Comparing with~\eqref{eq:Vepsdef} at small $T-T_c$, we identify
\be
  \eps = \frac{6s_c(T-T_c)}{c q_*} \ . 
\ee
The difference between this relation and~\eqref{eq:epsoptA} is not a surprise because the physical interpretation of the parameters $c$ and $q_*$ is also different between the two options. 

Because we had to make a specific choice for $\Gamma(\gamma)$ in equation~\eqref{e.GammaoptB} for the effective potential to have the desired form for option~B, this option may seem less preferable than option~A. We however remark that the same choice of $\Gamma(\gamma)$ gave an extremely good fit of the numerical data for the static domain wall solution. Therefore there is in practice very little freedom in choosing the function, and it is interesting that the same function works both for the accelerating solution and the numerical data.

In summary, we have shown that the coupling between $\gamma$ and hydrodynamics automatically reproduces the accelerating solution~\eqref{eq:gammatcov} without the need of introducing an asymmetry in the potential $V(\gamma)$ explicitly. This observation strongly supports our model. The accelerating solutions are however found with one potentially important limitation: above we assumed implicitly that the temperature $T$ is independent of the coordinates $x$ and $t$. While this is reasonable for the static solution, the backreaction of the accelerating solution to the hydrodynamics is expected to lead to a nontrivial profile to the temperature. That is, the simple accelerating solutions are only found when the backreaction to the hydrodynamics due to the wall motion is neglected.

The success of the model with the accelerating solution raises an important question which we already touched upon above: to which extent can we use the accelerating solutions to pin down the functions in the model, in particular the choices of $\Gamma(\gamma)$? It is clear that the solution~\eqref{eq:gammatcov} is only (at least up to trivial modifications) obtained with the choices of $\Gamma(\gamma)$ we did above, for both options~A and~B. However, the physical solution does not need to have exactly the form of~\eqref{eq:gammatcov}. For example, the shape of the profile of the domain wall could receive nontrivial corrections in the accelerating case. In any case we should require that the solution for $\gamma(x,t)$ continues to asymptote to zero and one in the deconfined and confined phases, respectively. While this is in principle a matter of (re)definition of $\gamma$ and $\Gamma(\gamma)$, possible variations would need to be compensated by modifying the coupling between $\gamma$ and the hydrodynamic degrees of freedom since the action for them far from the domain wall should be unchanged. This would lead to a complicated action which we do not want. Therefore the function $\Gamma(\gamma)$ should match with the forms used above at least near $\gamma = 0$ and $\gamma = 1$. For other values of $\gamma$, there is in principle no constraint, but apparently our choices are the only ones leading to potentials $V_\mathrm{eff}$ that are low-order polynomials. In any case the choice for $\Gamma(\gamma)$ must be quite close to those used here, otherwise the agreement with the numerical solutions found in figure~\ref{fig:Tmunufinal} is spoiled.
 
Finally, let us comment on the implications for the temperature dependence of the potentials $a$ and $V$ as well as the resulting energy-momentum tensor (which we already discussed above in section~\ref{sec:gammaaction}). Notice that, as we have demonstrated, accelerating solutions do not require any explicit temperature dependence in these functions. However we wish to check if such a dependence can be added without damaging the nice picture found above. 

Since the coupling to hydrodynamics automatically introduces the needed asymmetry in the effective potential of $\gamma$ for $\eps \ne 0$, it makes no sense to introduce an asymmetry in $V(\gamma,T)$ explicitly (even though it would be possible to do this). Modifying the overall coefficient of the potential, in the same way as in~\eqref{e.aV}, but with constant $\beta$, is however possible. To lowest order in $\eps$ this simply means renormalizing the value of the coupling in the potential, i.e., $q_* \to q_*(1+c_q \eps)$, everywhere. Similarly, one can introduce temperature dependence of the overall coefficient of $a(\gamma,T)$. 
But actually this function is even less constrained by the accelerating solutions. This happens because, as we noticed in section~\ref{sec:gammaaction}, the static $T=T_c$ $tanh$ profile satisfies the equations of motion from the action for $\gamma$ in~\eqref{eq:gammaaction} for any $a(\gamma)$. It follows that the $\mathcal{O}(\eps)$ domain wall solution only depends on the leading $\mathcal{O}(\eps^0)$  term of $a(\gamma,T)$. Therefore we can have
\be
 a(\gamma,T) = a(\gamma) + \eps \hat a(\gamma) + \mathcal{O}\left(\eps^2\right)
\ee 
with any $\hat a(\gamma)$ without affecting the  accelerating solution at $\mathcal{O}(\eps)$. The coefficient $\alpha$ in~\eqref{e.aV} is $\alpha \propto \hat a(\gamma)/a(\gamma)$ in the limit $T \to T_c$. Putting these observations together, the $\gamma$ dependence required for the formula~\eqref{e.Balphabeta} to agree with our choice $B=1+\Gamma$ can arise from $a(\gamma,T)$ through the coefficient $\alpha$.

\section{Comments on generality}
\label{s.generality}

The description proposed in the present paper was developed based on a conformal\footnote{Note that conformality refers only to the bulk and the physical boundary theory dimensionally reduced on the $\phi$ circle is of course nonconformal.} version of the holographic Witten's model~\cite{Witten:1998zw} for which there exists the numerical solution of the domain wall~\cite{Aharony:2005bm} interporating  between coexisting confining and deconfined phases. We believe, however, that the resulting framework should have a much more general range of applicability.

Indeed quite different (nonconformal) holographic models, which have a $1^{st}$ order phase transition between two \emph{deconfined} phases exhibit a domain wall profile which is very well fitted by a $tanh$ function~\cite{Attems:2019yqn,Attems:2020qkg}.
Since an underlying $tanh$ shape of the domain wall arises analytically in our framework, this suggests that the framework could also be applied in those settings. In appendix~\ref{s.nonconformal} we checked that indeed our description works quite well also for domain walls in the nonconformal bottom-up holographic model of~\cite{JJS} with two deconfined phases for which we had numerical data.

The reasoning leading to our proposal was in fact very general and the insight from holography was essentially used \emph{technically} only to constrain the form of $\Gm(\gm)$, $a(\gm)$ and $B(\gm)$ and eliminate possible nondiagonal terms\footnote{Such terms could nevertheless still reappear if they would be multiplied by $(u\cdot v)$.} $u^\mu v^\nu$. The key qualitative input from holography was really the reassurance that such a simple model could indeed work so well even for a strongly interacting field theory with the very nontrivial gravitational configuration dual to the domain wall~\cite{Aharony:2005bm}.

For a general system with a $1^{st}$ order phase transition we thus expect that one could model the energy-momentum tensor as
\eq
\label{e.generalized}
T_{\mu\nu} = (1-\Gm)\, T_{\mu\nu}^{phase\;A} + \Gm\, T_{\mu\nu}^{phase\;B} + T_{\mu\nu}^\Sg
\eqx
with $\Gm$ being a function of $\gm$ and the surface tension energy momentum tensor $T_{\mu\nu}^\Sg$ following from an action of the form
\eq
\LL_\gm = - a(\gm, T) \left(  \f{1}{2} (\partial \gm)^2 +V(\gm, T) \right)
\eqx
with
\eq
V(\gm, T) \propto \gm^2 (1-\gm)^2\ .
\eqx
With the current available data we cannot differentiate between options A and B or rule out an alternative, but the two choices $a(\gm,T)= c(T)/\gm(1-\gm)$ and $a(\gm,T)=\tilde{c}(T)$ seem most natural.

Let us note finally an interesting interpretation of our framework (applicable only in case of option B i.e. with the canonical scalar action with $a(\gm)=1$). Consider the overall action density for option B
\eq
\label{e.overallactionB}
\LL = (1-\Gm(\gm))p(T) + \Gm(\gm) -\f{1}{2} c(T) (\partial \gm)^2 - d(T) \gm^2 (1-\gm^2)
\eqx
where we explicitly separated out temperature dependence and emphasised that $\Gm$ is a known functional\footnote{For the system investigated here it is given by $\Gm(\gm)=\gm^2 (3-2\gm)$.} of $\gm$, which is the elementary field here. The Euclidean Lagrangian aims to model the holographic on-shell action\footnote{This was the reason that we used the hydrodynamic action formulation of~\cite{Mukund} instead of~\cite{Hydroaction}.} which evaluates holographically the free energy. So restricting (\ref{e.overallactionB}) to constant configurations gives an expression for the free energy as a quartic polynomial in $\gm$:
\eq
\label{e.Fquartic}
F = -(1-\Gm(\gm))p(T) - \Gm(\gm) + d(T) \gm^2 (1-\gm^2)
\eqx
which essentially provides a Landau type description of a $1^{st}$ order phase transition. Our formulation (\ref{e.overallactionB}) can be then understood as promoting the Landau order parameter $\gm$ to a dynamical effective field and coupling it in a natural way with hydrodynamic degrees of freedom.
The above interpretation fails for option A due to the $1/\gm (1-\gm)$ prefactor of the scalar field action and thus also of the potential.
So from this point of view option B might be preferable.

\section{Summary and outlook}

In the present paper we investigated theories with a $1^{st}$ order phase transitions with special emphasis on the description of domain walls separating regions of coexisting phases.
A~striking feature of the dual holographic descriptions of domain walls in various theories is that the domain wall profiles appear to be much simpler than one would expect from the rather complicated gravitational backgrounds. This is especially apparent for the case of Witten theory, where the plasma-ball domain wall gravitational solution~\cite{Aharony:2005bm} is particularly nontrivial.

The observation that the boundary physics seems much simpler than the full gravitational description, 
suggests that one could model it directly on the level of the boundary field theory energy-momentum tensor. To this end, we introduced an additional degree of freedom $\gm$, akin to an order parameter distinguishing the two phases and proposed a unified description incorporating both the energy-momentum tensors of the two coexisting phases as well as a~piece localized essentially in the vicinity of the domain wall and describing its surface tension profile.

In the case of the Witten model~\cite{Witten:1998zw} with a confining and deconfined phases, we arrived at a description in terms of a simple action for $\gm$ with a quartic potential coupled with hydrodynamics of the plasma in the deconfined phase.
For the static planar domain wall, one can obtain simple analytical formulas for all components of the energy-momentum tensor.

We believe that the overall framework should be applicable in a very general context of theories with a $1^{st}$ order phase transition.
We checked that indeed one can get a very good description also for the Witten model in another dimensionality as well as for a nonconformal bottom-up holographic model from~\cite{JJS} with two coexisting plasma phases, but we expect that similar modelling should hold even beyond holography.

The obtained results lead to a wide variety of open questions which should spark various directions of further research.
\begin{itemize}
\item
The holographic data at our disposal does not allow us to distinguish between two variants of the proposed action (option A and option B), as well as to control the variations of the parameters as we move away from $T=T_c$. Even at $T=T_c$, we find that the $B(\Gm)$ coefficient in the domain wall energy-momentum tensor depends on the specific theory.
Indeed, one can view these parameters as analogs of transport coefficients in hydrodynamics, hence an understanding of their variability in different holographic models would be very interesting.
\item
Going further along these lines, it would be interesting to investigate natural ways of systematically improving the model by including e.g. higher derivative terms in an analogous way to going to higher order hydrodynamics or refining the scalar potential.
\item
An obvious question that raises from the model proposed in this paper is how to interpret the field $\gamma$ as an order parameter. In particular, 
  $1-\gamma$  as an order parameter for deconfinement. One may  try to relate it to the standard order parameters for deconfinement, like the VEV of a Polyakov loop which in principle one can try to compute holographically as a function of $x$ in the domain wall background, and see if its profile is similar to that found for   $\gamma$\footnote{We thank Ofer Aharony for raising this point}.   
\item
It would be  very important to have a wide range of holographic gravitational solutions which would allow one to clarify these issues. Time dependent solutions of the holographic Einstein equations would be especially instructive, as these could fix possible $(u\cdot v) \left(u^\mu v^\nu + v^\mu u^\nu -\f{1}{2} (u\cdot v) \eta^{\mu\nu}\right)$ terms in~$T_{\mu\nu}$ and corresponding coupling terms in the scalar action for $\gm$. As a first step, it should be instructive to 
construct the gravity solutions for a spherical domain wall~\cite{Figueras:2014lka}, and for a flat accelerating one.

\item
Going further along these lines, one could investigate the incorporation of dissipation within the proposed framework.
Indeed understanding the impact of dissipation in the above context should be quite important for the relevant physics. 
\item
The whole processes of transition from one phase to another in both directions are quite  complex and are believed to include bubble nucleation and bubble coalescing. It will be interesting to analyze these sub-processes using the action  and the energy momentum tensor proposed in this paper.
\item 
The model proposed in this paper has been inspired by holography. Thus it is related a priori to gauge dynamical systems with large $N_c$ and large $\lambda_\mathrm{'t\, Hooft}$. A very interesting  question is to what extend can one extrapolate from the results derived here also to real world of $N_c=3$ and $\lambda_\mathrm{'t\, Hooft}\sim 1$. One obvious question is the nature of the transition.
\item
Another obvious question that follows the model proposed in this paper  is can one determine using it physical properties that can be experimentally observed. 
Using the hydrodynamical description of the quark gluon fluid  such properties were computed and compared to experimental data.  The question is what additional information can be extracted about them using our domain-wall model.
\item
The action of our model resemble the one in  Coleman's seminal papers on  bounce~\cite{Coleman:1977py, Callan:1977pt}. It would be interesting to use the results of the latter papers to   determine  the decay width of the ``false vacuum"  for systems at temperature different from the critical one.
\item 
While the covariant description presented in this article was strongly motivated by numerical data obtained through holography, we did not derive any of the ingredients of the model by using the gravity dual. Naturally, such a derivation would  be desirable. An attempt in this direction was presented in section~\ref{s.toymodel}, where we studied a simple model for the domain wall geometry in the Witten's model. This model however failed to explain some features of the numerically obtained exact geometry, and we found no way to improve it systematically towards an exact solution of the Einstein equations. It would be extremely interesting if the connection between the gravity solutions and the boundary description could be made more precise either by improving this attempt or by some other means.
\item
Since finding domain wall solutions to Einstein equations is typically rather difficult, it may be useful to explore ways to simplify the problem on the gravity side. Such a simplifying approach might be to use the limit of large number of dimensions. This is actually well motivated for domain walls or phase transitions in QCD: as was pointed out in~\cite{Betzios:2017dol,Betzios:2018kwn}, a class of geometries which produces IR features similar to QCD~\cite{Gursoy:2008za} is well approximated by a dimensional reduction of a high dimensional AdS space.  
\item
In this paper we have considered the holographic description of glue dynamics. An obvious question is how would the incorporation of fundamental quarks affect the system. It is well known that such a system is characterized in addition to the confining/deconfining transition also the one associated with chiral symmetry breaking.
The holographic dual   of both phase transitions, which incorporates flavor branes was analyzed in \cite{Aharony:2006da}. It will be very interesting to construct a model analogous to the one presented here that includes also quark degrees of freedom and correspondingly mesons and baryons. 
\item
As was mentioned above the model proposed in this paper has been inspired by holographic duality with a gravitational background. However, the holographic duality is in fact an even deeper type of duality than the form in which it is commonly used. It relates the gauge dynamics of a boundary field theory and \emph{string theory} that resides in the bulk. In \cite{Sonnenschein:2016pim, Sonnenschein:2014jwa} it was argued that this type of duality does a much better job in describing the spectra and decay processes of hadrons. Thus, the confining phase should be described by string degrees of freedom.  It would be interesting to uplift the model of this paper into a stringy model which may yield a better description of the hadronization process.
\end{itemize}

\bigskip
\noindent{}{\bf Acknowledgements.} We would like to thank Maximilian Attems, Tuna Demircik, Wojciech Florkowski, Mukund Rangamani, Toby Wiseman for interesting discussions. We would also like to thank Ofer Aharony for his remarks on the manuscript. The numerical data from~\cite{Aharony:2005bm} was reproduced by using a computer program written by Toby Wiseman. 
The  work 
 of JS supported in part by a center of excellence supported by the Israel Science Foundation (grant number 2289/18).
The research of MJ is supported by an appointment to the JRG Program at the APCTP through the Science and Technology Promotion Fund and
Lottery Fund of the Korean Government. The research of MJ is also supported by the
Korean Local Governments -- Gyeongsangbuk-do Province and Pohang City, and by the National Research Foundation of Korea (NRF) funded by the
Korean government (MSIT) (grant number 2021R1A2C1010834).

\appendix

\section{The domain wall in a six-dimensional setup}\label{app:d4}

In this appendix we discuss the results of section~\ref{s.phenomenology} for a generalization of the domain wall geometry which has one additional spatial dimension, and was also discussed in~\cite{Aharony:2005bm}. One can consider even more general solutions of the Einstein equations with a cosmological constant for $d+2$-dimensional gravity, with $d$ equalling the number of spatial dimensions (including the compactified dimension), but we restrict here to $d=3$ (the main text) and $d=4$ (this appendix). Notice that the six-dimensional geometry therefore differs from the Witten's model~\cite{Witten:1998zw} of compact  D4 background~\cite{Brandhuber:1998er} as there is no dilaton (or the dilaton is constant) and the background is conformal. That is, the geometry is closely related to AdS$_6$: in particular, in the deconfining and confining limits it is obtained from~\eqref{eq:deconfgeom} and~\eqref{eq:confgeom}, respectively, simply by adding one spatial dimension (i.e., now $i,j = 1 \ldots 3$ in these formulas) and modifying the blackening factor to
\be
 f(u) = 1 - \left(\frac{4 \pi R T_c}{5u}\right)^{5} \ ,
\ee
where $R$ is the radius of the AdS$_6$ space.

\begin{figure}[t!]
\begin{center}
\includegraphics[width=.47\textwidth]{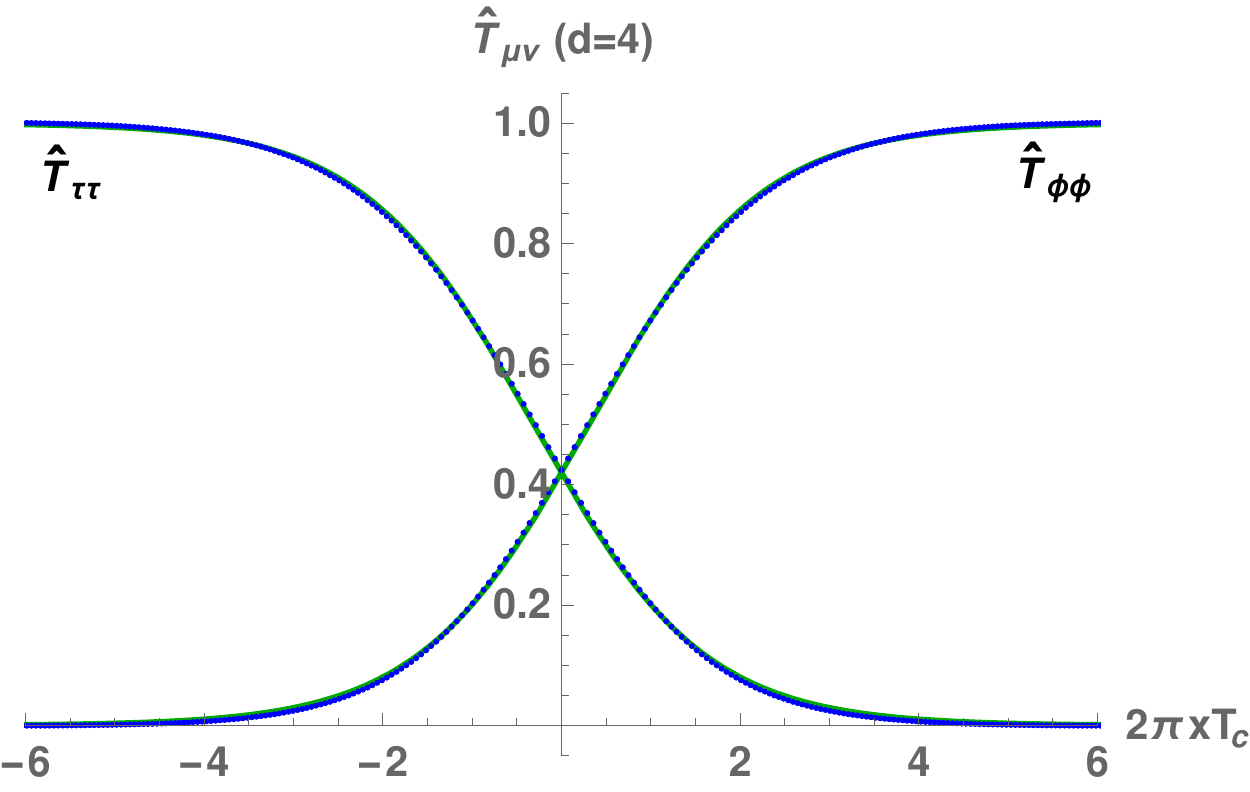}%
\hspace{4mm}
\includegraphics[width=.47\textwidth]{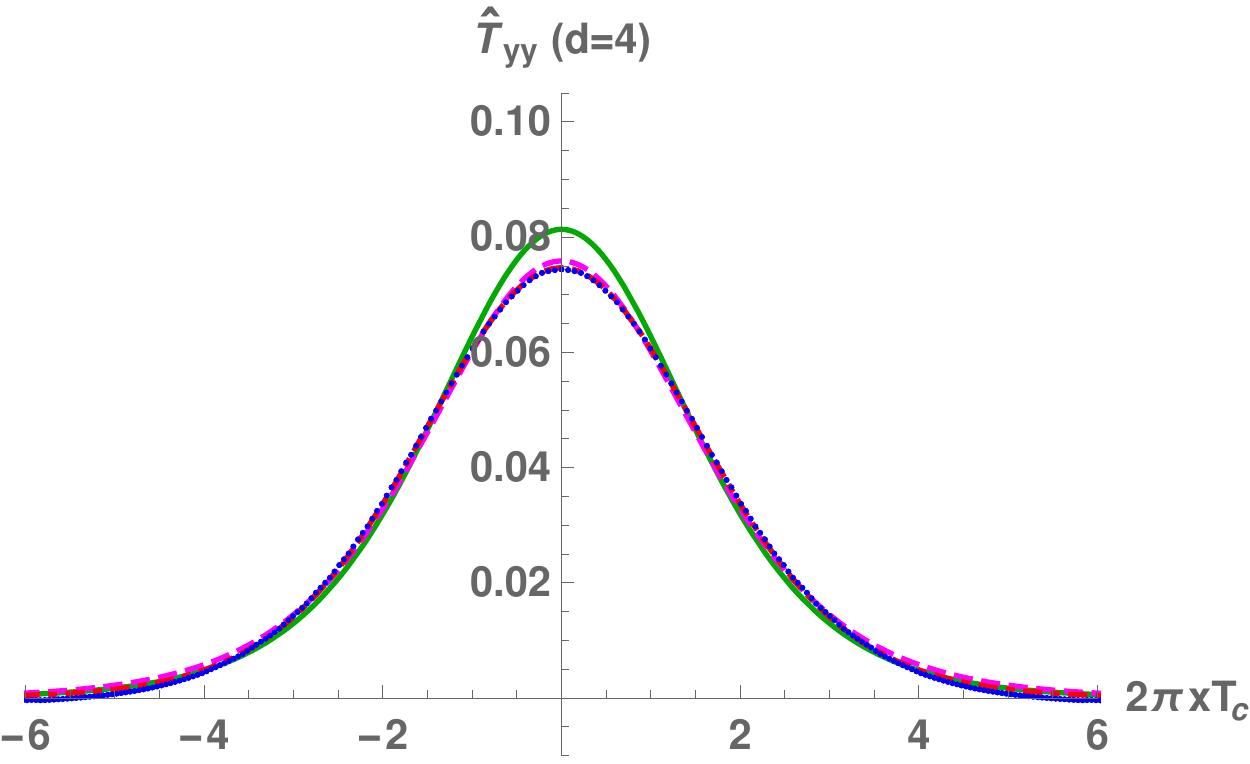}%
\end{center}
\caption{Fitting of the (normalized) components of the energy-momentum tensor for the six-dimensional geometry. 
Left: $tanh$ fit to $\hat T_{\tau\tau}$ and to $\hat T_{\phi\phi}$.
Blue dots show the energy-momentum tensor extracted from the numerically constructed geometry of~\protect\cite{Aharony:2005bm}, and the green curves are given by the fit in equation~\protect\eqref{eq:tanhfit}. Right: Fits to $\hat T_{yy}$.  The green curve is the prediction from the fit of figure~\protect\ref{fig:TttandTphiphi}, given in equation~\protect\eqref{eq:Tyypred}. The dashed magenta and dot-dashed red fits were obtained by using the former and latter formulas of equation~\protect\eqref{eq:coshfit}, respectively. 
\label{fig:Tmunud4}}
\end{figure}

We may extract the components of $T_{\mu\nu}$ from the domain wall solution and fit them by using the Ans\"atze in~\eqref{eq:tanhfit} and in~\eqref{eq:coshfit} in the same way as for the five-dimensional geometry in section~\ref{s.phenomenology}. The results are shown in figure~\ref{fig:Tmunud4}, where the left (right) plot is the counterpart of figure~\ref{fig:TttandTphiphi} (figure~\ref{fig:Tyy}). The notation is as in section~\ref{s.phenomenology}. The fit parameters for the $tanh$ fit are given by
\begin{align} 
\frac{q_*}{2\pi T_c} & \approx 1.054 \ , \qquad 2 \pi T_c x_0 \approx 0.311 \ , & &(d=4)& \label{eq:tfitresd4} 
\end{align}
for the $tanh$ fit and by
\begin{align} 
    \frac{q_*^{(2)}}{2\pi T_c} & \approx 0.981 \ , \qquad c^{(2)} \approx 0.0758 \ , & & & \nonumber\\
    \frac{q_*^{(4)}}{2\pi T_c} & \approx 0.652 \ , \qquad c^{(4)} \approx 0.0747 \ , & &(d=4)\ ,& 
\end{align}
for the $cosh$ fits. 

The main observation from these fits is that dependence on $d$ appears to be mild. The ratios between the various values of $q_*$ are essentially unchanged. The value of $x_0$ is a bit higher than for $d=3$, but $\hat T_{yy}$ is in turn smaller. This happens because~\eqref{eq:Tyyrel} depends on $d$ and reads, in general, 
\be \label{eq:Tyyrelgen}
 \hat T_{yy}(x) = \frac{1}{d-2}\left(1-\hat T_{\tau\tau}(x)-\hat T_{\phi\phi}(x)\right) \ .
\ee

\begin{figure}[t!]
\begin{center}
\includegraphics[width=.8\textwidth]{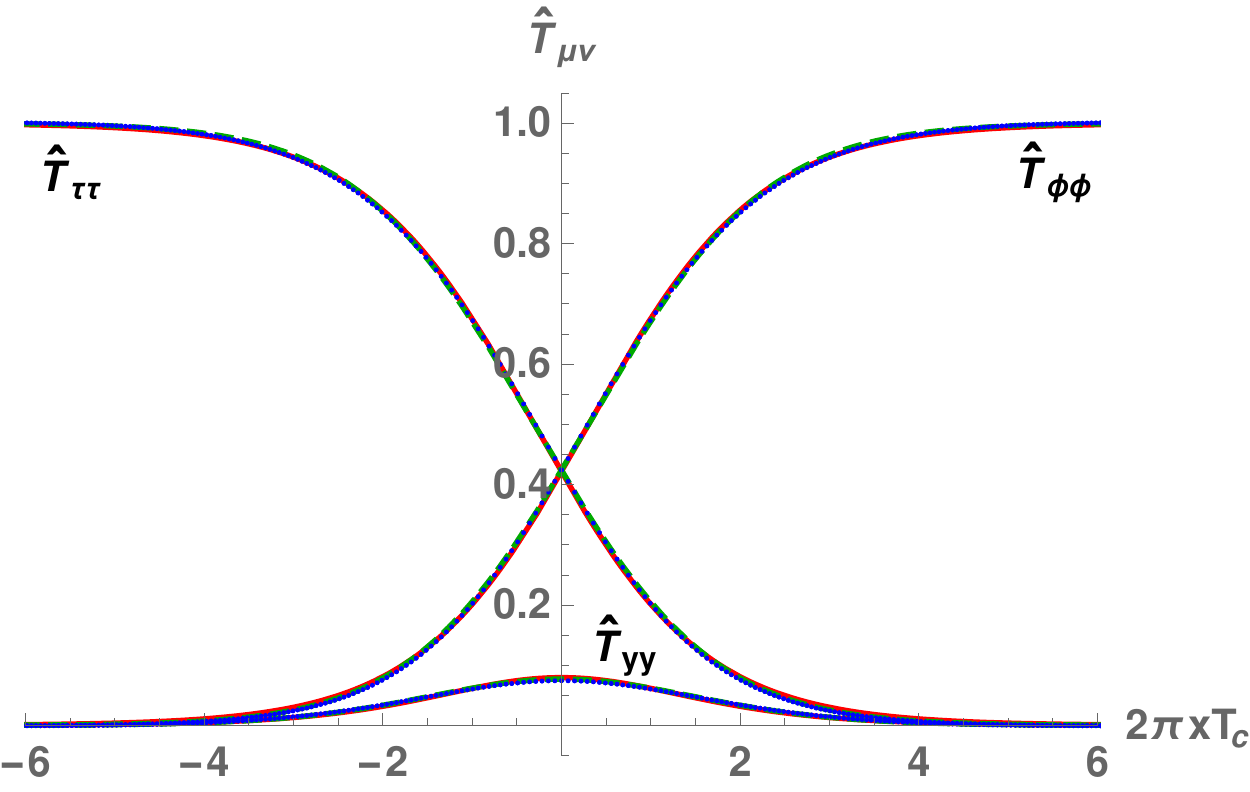}
\end{center}
\caption{Final fits of the model to the numerical solution for the domain wall for $d=4$. The blue dots, solid red curve, and dashed green curve show the numerical data, fit using the option A, and fit using the option B, respectively. \label{fig:Tmunufinald4}}
\end{figure}

We have also carried out the fit to the data by using our final model for the energy momentum tensor in~\eqref{e.tmunutotal}\footnote{For $d=4$ the coefficient of the $n_\mu n_\nu$ term is $\Sigma(4-B)$.}. As it turns out, for $d=4$ a good fit to data is obtained by using
\be
 B(\Gamma) = 2
\ee
instead of $B(\Gamma)=1+\Gamma$, which was used for $d=3$ in section~\ref{s.tmunufits}. The fit results for $d=4$ are shown in figure~\ref{fig:Tmunufinald4} and is almost identical to the result for $d=3$ in figure~\ref{fig:Tmunufinal}. That is, the fit is again excellent. The deviation of the data from the fit (which is poorly visible in the plots) is of the same magnitude in all fits, but a bit larger for $d=4$ than for $d=3$.

\section{The domain wall for a nonconformal model with two deconfined phases}
\label{s.nonconformal}

\begin{figure}[th]
\begin{center}
\includegraphics[width=.47\textwidth]{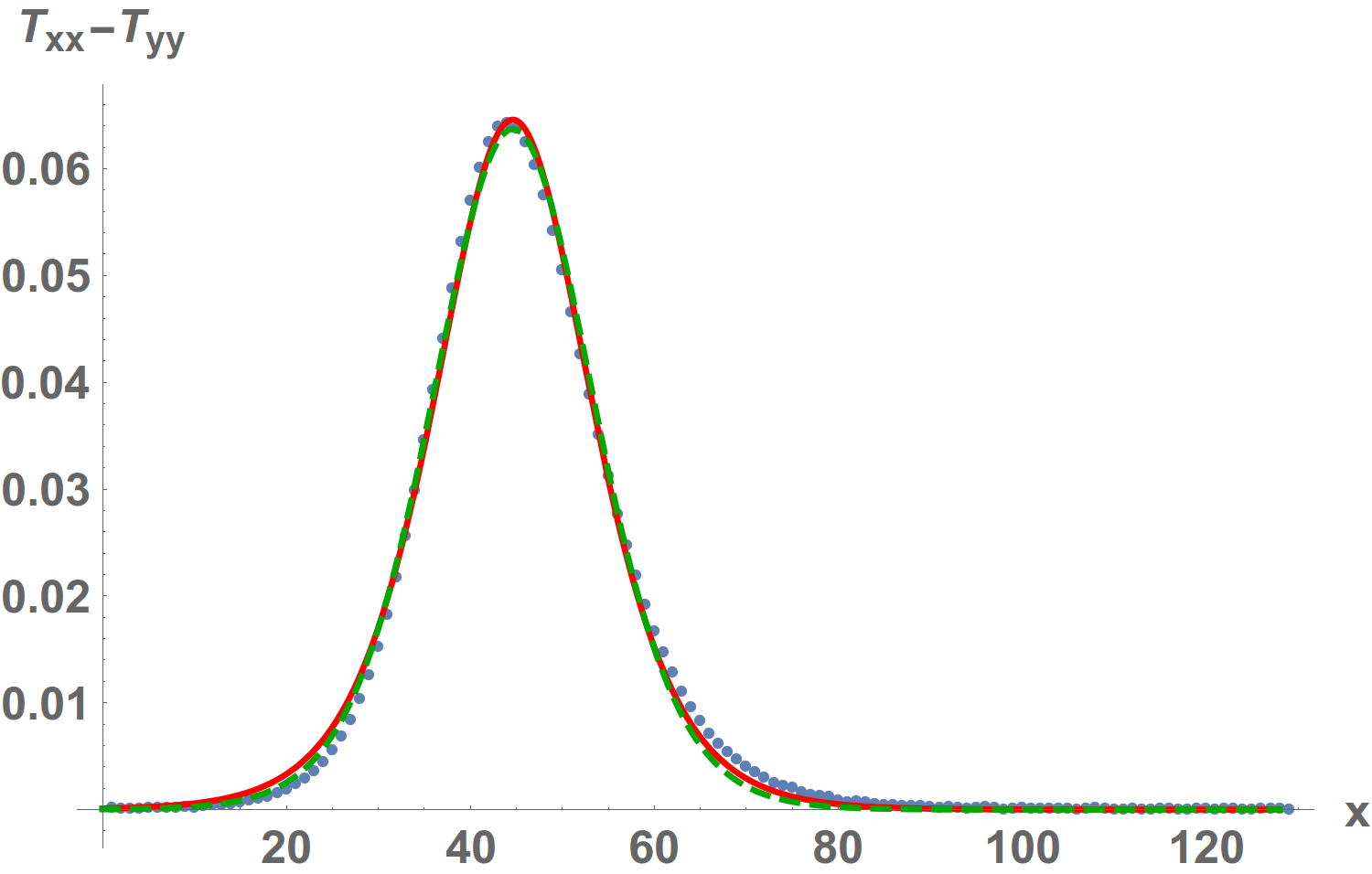}%
\hspace{4mm}
\includegraphics[width=.47\textwidth]{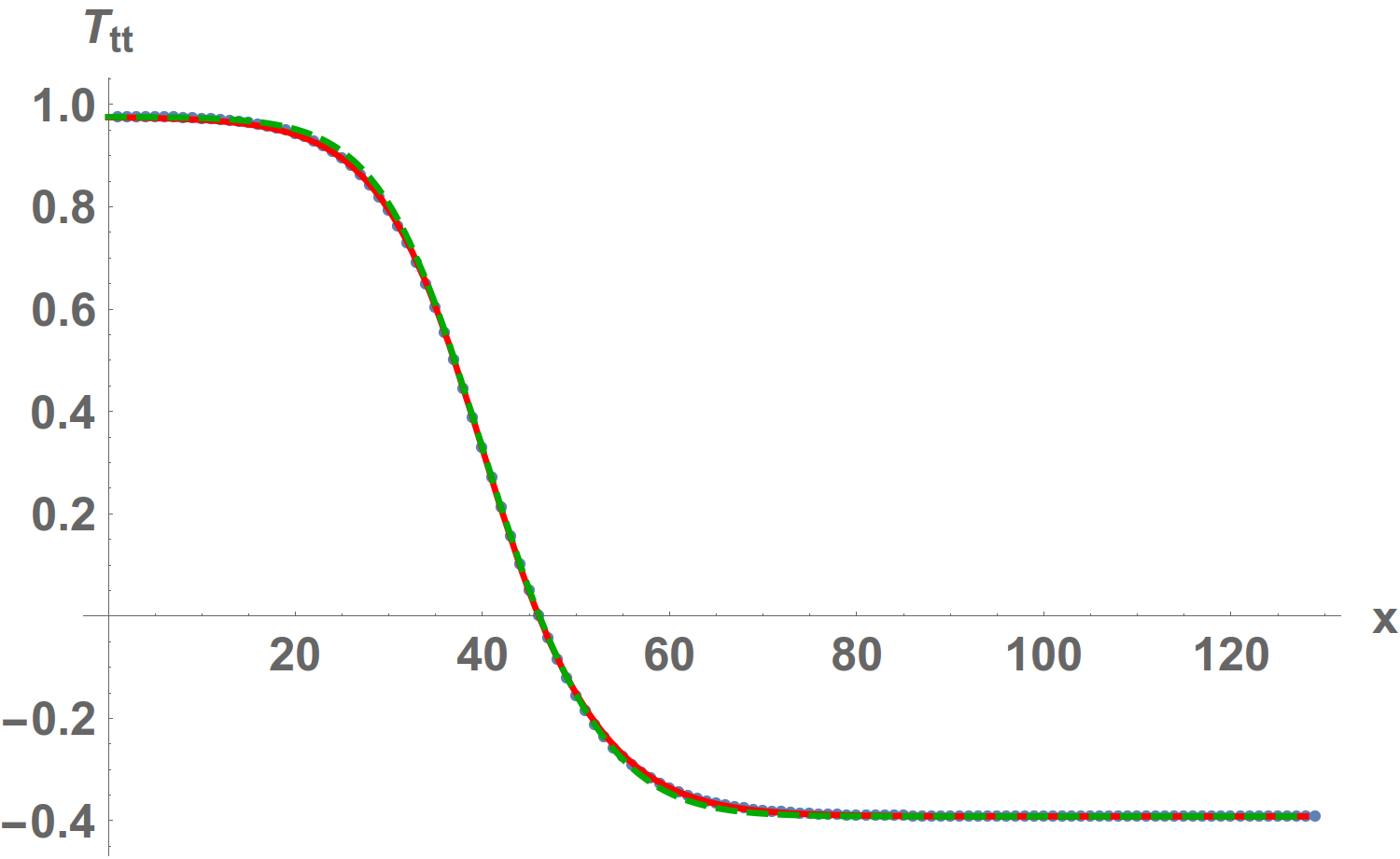}%
\end{center}
\caption{Reproducing the numerical domain wall between two deconfined phases for the nonconformal bottom-up holographic gravity+scalar theory from~\cite{JJS}.
On the left: surface tension density $T_{xx}-T_{yy}$ fitted using option  A (red) and option B (dashed green). On the right: energy density $T_{tt}$ using the same $B(\Gm)$ for both options. The numerical data marked by blue dots are taken from simulations performed in~\cite{JJS}.
\label{fig.nonconformal}}
\end{figure}

In this appendix we check that our framework is applicable in the context of a quite different holographic model studied in~\cite{JJS}
-- a bottom-up gravity+scalar system in a 4D bulk with the scalar field potential 
\eq
V(\Phi) = -6 \cosh \left(\f{\Phi}{\sqrt{3}}\right) -0.2\, \Phi^4
\eqx
The scale breaking conformality is set by a nonzero source for the operator dual to the scalar field $\Phi$.
This model exhibits a $1^{st}$ order phase transition with the two coexisting phases being deconfined, both described holographically by black holes. The numerical simulations in~\cite{JJS} followed the evolution from the spinodal instability to the formation of domains of the two coexisting phases separated by domain walls. Here we analyze one of the domain walls appearing in the final state configuration.

We first fit the $q_*$ and $c$ parameters for options A and B  (recall (\ref{e.optionA})-(\ref{e.gammaoptions})) from the surface tension density $\Sg(x) = T_{xx}-T_{yy}$. The outcome is shown in figure~\ref{fig.nonconformal} (left).
Consequently we have
\eq
\gm(x) =\f{1}{2} + \f{1}{2} \tanh  \f{q_* (x-x_{center})}{2} 
\eqx
where $x_{center}$ is the position of the center of the domain wall, which was not \emph{a-priori} fixed in the numerical simulation of~\cite{JJS}.
We then determine $B(\Gm)$ by reproducing the energy density $T_{tt}$ using the formula
\eq
T_{tt} = (1-\Gm)\, \epsilon_{phase\;A} + \Gm\, \epsilon_{phase\;B} +(1-B(\Gm))\,\Sg
\eqx
where we use the generalized version (\ref{e.generalized}) which takes into account appropriate energy densities of the two distinct deconfined phases.
The last term follows from
\eq
T_{\mu\nu}^\Sg = \Sg \left(
- \eta_{\mu\nu} +  v_\mu v_\nu -B u_\mu u_\nu \right)
\eqx
which differs from (\ref{e.tmunusigma}) just by the absence of the $n_\mu n_\nu$ term as in the present case there is no $\phi$ direction.
Recall that for option A, $\Gm(\gm)=\gm$, while for option B, $\Gm(\gm)=\gm^2 (3-2\gm)$. We use a common $B(\Gm)$ in both cases. We find that for this theory
\eq
B(\Gm) = 6.25 -3.5\, \Gm
\eqx
works very well, as shown in figure~\ref{fig.nonconformal} (right).

\section{The $n^\mu n^\nu$ terms from an action}
\label{s.nmunnu}

For completeness, let us discuss how one could formally implement the $n^\mu n^\nu$ in the energy-momentum tensor as coming from an action.
Now of course the action would have to be considered as a \emph{four-dimensional} action including the $\phi$ circle.
We should just look at solutions which do not involve any dependence on $\phi$. 

A natural way to proceed is to introduce a vector $N^\mu$ and
\eq
N = \f{1}{\sqrt{g_{\mu\nu} N^\mu N^\nu}}
\eqx
so that the unit vector $n^\mu$ is given by
\eq
n^\mu = N N^\mu \ .
\eqx 
All this is rather artificial as we have to keep $N=1$ and $N^\mu=(0,0,0,1)$ fixed and nondynamical. Proceeding as before, we get the variation
\eq
\dl N = \f{1}{2} N n_\mu n_\nu \dl g^{\mu\nu}
\eqx
and we see that the confining energy-momentum tensor 
\eq
T_{\mu\nu}^{conf} = g_{\mu\nu} -4 n_\mu n_\nu
\eqx
arises from the Lagrangian density
\eq
\LL^{conf} = q(N) = N^4 \ .
\eqx

Let us now move to the scalar action. The only change will be that the coefficients can now be functions of both $T$ and $N$. Again for simplicity we take a power law Ansatz
\eq
a(T,N,\gm) =  T^\al N^\kappa a(\gm) \ ,\qqqq
V(T,N,\gm) =  T^\bt N^\rho V(\gm) \ .
\eqx
In the formula for $T_{\mu\nu}^\Sg$ we thus have the following terms (evaluated at $T=N=1$)
\eq
-a(\gm)\left(\alpha (\partial \gm)^2 + (\alpha+\beta) V(\gm) \right) u_\mu u_\nu + a(\gm)\left(\kappa (\partial \gm)^2 + (\kappa+\rho) V(\gm) \right) n_\mu n_\nu
\eqx
as well as 
\eq
 a(\gm) \partial_\mu\gm \partial_\nu\gm - a(\gm) \left( \f{1}{2} (\partial \gm)^2 +V(\gm) \right) g_{\mu\nu}\ .
\eqx
Now we should impose the tracelessness condition. We can do it now as currently we have a model of the energy-momentum tensor of the \emph{four-dimensional} boundary theory. We get two equations arising from the coefficients of $(\partial \gm)^2$ and $V(\gm)$ in the trace:
\eq
\alpha + \kappa = 1 \, \qqqq \alpha + \beta +\kappa +\rho = 4\ .
\eqx
The numerical fits lead to
\eq
B = \f{3}{2} \alpha + \f{1}{2} \beta = 1 + \Gm \ , \qqqq
-C = \f{3}{2} \kappa + \f{1}{2} \rho = 2 - \Gm
\eqx
where $B$ and $C$ are the coefficients in (\ref{e.tmunusigma}).
But again we would like to emphasize that this is just a formal  exercise as we always take $N\equiv 1$ in the Witten model~\cite{Witten:1998zw}.

\end{document}